\documentclass[aps,prc,superscriptaddress,showpacs,onecolumn]{revtex4}	
 \usepackage{amsmath}  
 \usepackage{makeidx}
 \usepackage{graphicx}
 \usepackage{amsfonts}
 \usepackage[ansinew]{inputenc}
 \usepackage[usenames,dvipsnames]{pstricks}
 \usepackage{subfigure}
 \usepackage{epsfig}
 \usepackage{pst-grad}
 \usepackage{pst-plot}
 \usepackage{mathrsfs}

              \makeindex
              
\begin{document}
\title{Bound clusters on top of doubly magic nuclei}
\author{G. R\"{o}pke}
\email{gerd.roepke@uni-rostock.de}
\affiliation {Institut f\"{u}r Physik, Universit\"{a}t Rostock, D-18051 Rostock, Germany}
\author{P. Schuck}
\affiliation{Institut de Physique Nucl\'{e}aire, Universit\'e Paris-Sud, IN2P3-CNRS, UMR 8608, F-91406, Orsay, France}
\affiliation{Laboratoire de Physique et Mod\'elisation des Milieux Condens\'es, CNRS- UMR 5493, 
F-38042 Grenoble Cedex 9, France}
\author{Bo Zhou}
\affiliation{Department of Physics, Nanjing University, Nanjing 210093, China}
\affiliation{Research Center for Nuclear Physics (RCNP), Osaka University, Osaka 567-0047, Japan}
 \author{Y. Funaki}
 \email{funaki@riken.jp.}
\affiliation{ Nishina Center for Accelerator-Based Science, The institute of Physical and Chemical Research (RIKEN), 
Wako 351-0198, Japan}
 \author{H. Horiuchi}
 \affiliation {Research Center for Nuclear Physics (RCNP), Osaka University, Osaka  567-0047, Japan}
 \affiliation {International Institute for Advanced Studies, Kizugawa 619-0225,  Japan}	
\author{Zhongzhou Ren}
\affiliation{Department of Physics, Nanjing University, Nanjing 210093, China}
\affiliation{Center of Theoretical Nuclear Physics, National Laboratory of Heavy-Ion Accelerator, Lanzhou 730000, China}
 \author{A. Tohsaki}
 \affiliation{Research Center for Nuclear Physics (RCNP), Osaka University, Osaka 567-0047, Japan}
\author{Chang Xu}
\affiliation{Department of Physics, Nanjing University, Nanjing 210093, China}
\author{T. Yamada}
\affiliation{Laboratory of Physics, Kanto Gakuin University, Yokohama 236-8501, Japan}

\date{\today}

\begin{abstract}

An effective $\alpha$ particle equation is derived for cases where an $\alpha$ particle is formed on top of a doubly magic nucleus. 
As an example, we consider $^{212}$Po with the $\alpha$ on top of the $^{208}$ Pb core. 
We will consider the core nucleus infinitely heavy, so that the $\alpha$ particle moves with respect to a fixed center, i.e., recoil effects are neglected. The fully quantal solution of the problem is discussed.
The approach is inspired by the THSR (Tohsaki-Horiuchi-Schuck-R\"{o}pke) wave function concept 
that has been successfully applied to light nuclei. 
Shell model calculations are improved by including four-particle ($\alpha$-like) correlations that are of relevance when the matter density becomes low. 
In the region where the $\alpha$-like cluster penetrates the core nucleus, the intrinsic bound state wave function transforms at a critical density into an unbound four-nucleon shell model state.
Exploratory calculations for $^{212}$Po are presented.
Such preformed cluster states are only hardly described by shell model calculations.
Reasons for
different physics behavior of an $\alpha$-like cluster with respect to a deuteron-like cluster are discussed. 
\end{abstract}


\maketitle

 \section{\label{sec:introduction}
 Introduction}
 
  The shell model of nuclei has been proven as a very successful concept describing properly 
many features of the structure of nuclei \cite{RS}. A mean-field potential is introduced defining single nucleon states
 that are populated up to a maximum energy that  is the chemical potential or 
the Fermi energy of the neutrons or protons, respectively. Pairing can be included in a mean-field approach using a 
Bogoliubov transformation among the single particle orbits.
 In general, the treatment of correlations is a difficult problem in a single-nucleon mean-field approach beyond the two particle case with pairing.
 However, cluster formation may occur in special situations, and the systematic treatment of correlations 
 beyond the mean-field theory is a great challenge in the actual treatment of nuclear structure
 \cite{Wildermuth,RSM,Ren87,RSSN,Boneu,Bo,Bo2,Varga,Pol,Lovas,Ren,Wyss,Xu}.

 The problem of cluster formation in or on a nucleus is that, besides for the deuteron cluster,  
heavier clusters like $t, \,\,^3$He, $\alpha$ are very difficult to handle technically 
if one wants to treat the relative motion of the cluster versus the core nucleus correctly.
In principle, this is a very complicated three-, four-, etc., body problem. 
The solution should join two limiting cases, the situation where the cluster is well inside the core nucleus and a shell-
model mean-field calculation can be performed (Hartree-Fock-Bogoliubov), 
and the limit of distant clusters. In the present work, we focus on four-particle ($\alpha$-like)
correlations. Because of spin- isospin degeneracy, such correlations are quite 
strong and of relevance in low-density nuclear systems \cite{Wildermuth,RSM,Ren87}.

A main ingredient is the introduction of a collective variable, describing the center of mass (c.o.m.)
motion of the considered cluster, and variables that describe the intrinsic motion. A suitable choice 
are Jacobian coordinates, see Sec. \ref{homogen} for the four-particle case.
The separation of an energy eigenstate $\Psi$ of the few-particle cluster into a contribution $\Phi({\bf R})$
 where ${\bf R}$ denotes the c.o.m. coordinate, and an intrinsic part depending only on relative coordinates, 
 is strict for a homogeneous system
because the total momentum $ {\bf P}$  is conserved.
This simple decomposition is not possible for finite systems such as nuclei considered here.
As shown in Sec. \ref{comSchr}, in the general case of inhomogeneous systems such as nuclei, the intrinsic wave function 
$\varphi^{\text{intr}}({\bf r}_i-{\bf r}_j,{\bf R})$ of the cluster (we focus on four-nucleon clusters) 
also depends  on the c.o.m. position ${\bf R}$.
 This is mainly due to the Pauli blocking that depends on the local nucleon density near ${\bf R}$.
The cluster ($\alpha$-like) nucleonic wave function in momentum space is blocked out  inside the Fermi sphere.
Also the global form of the four-particle wave function changes from a Gaussian-like shape at low densities 
where $\alpha$ particle-like bound states are formed, 
to a shape which corresponds to the wave function of four single nucleons found in shell model states, see Sec. \ref{expl} and \cite{RSSN}.

The introduction of the c.o.m. coordinate ${\bf R}$ as a new dynamical collective degree of freedom simplifies the treatment 
of correlated nuclear systems beyond the single quasiparticle approximation. 
Although the shell model gives a complete basis,
the relevant coordinates are obtained only at the cost of very high configurational mixing. 
As an example, the THSR (Tohsaki-Horiuchi-Schuck-R\"{o}pke) ansatz \cite{THSR} to describe the Hoyle 
state in $^{12}$C is very successful because after separating the c.o.m. motion of the $\alpha$-like clusters from their intrinsic 
motion, a simple form for the wave function can be given that describes the Hoyle state in excellent approximation.
In contrast to the Brink ansatz, the c.o.m. motion should be treated dynamically \cite{Boneu} with a full freedom of the wave
function, for instance, in what concerns its extension what avoids the superposition of
many states with $\alpha$ clusters fixed at different positions.

The separation of the c.o.m. motion is crucial to simplify the problem in the case of  $\alpha$-cluster formation. It has been 
applied, besides $^8$Be,  also to other systems, like $^{16}$O or $^{20}$Ne \cite{Bo,Bo2}. It is always of importance that not only the 
separation of the c.o.m. motion from the intrinsic motion is performed, but also that the Pauli blocking is respected  
by the full antisymmetrization of the nucleonic wave function.
Once these ingredients are taken into account, the description is reasonable even in the case of the deuteron. 
One may consider  $^6$Li =  $\alpha$ + d, $^{18}$F =  $^{16}$O + d, or $^{210}$Bi = $^{208}$Pb + d. 
However, as will be discussed in more detail below, there exists a crucial difference between a two body cluster 
and the $\alpha$-like cluster. 
Namely, as we have shown in previous works \cite{RSSN,Sogo}, a quartet ($\alpha$-particle) dissolves very fast 
as a function of increasing baryonic density and around a nuclear matter density $n_{B,\text{cluster}} \approx n_0/5$, 
with $n_0\approx 0.15$ fm$^{-3}$ the saturation density, the $\alpha$ particle as a well formed cluster
has disappeared. The deuteron is also dissolved as a bound state, but Cooper pairing remains also at high densities.  
On the other hand, as we know, standard pairing persists to much higher densities and even beyond $n_0$. 
Reasons for this difference between the pairing and quartetting cases are given below. We want to neglect the recoil of the core. 
Then a heavier nucleus 
like $^{212}$Po  is a better choice, and the separation of the c.o.m. motion refers only to the $\alpha$-like cluster. 
Note that a similar problem to separate different degrees of freedom arises also in other 
fields such as the Born-Oppenheimer approximation in electron-ion systems \cite{Gross}.

The treatment of correlations in nuclei with one $\alpha$  on  top of doubly magic nuclei such as $^{212}$Po has a long-standing tradition, 
see \cite{Varga,Pol,Lovas,Ren,Wyss,Xu}. One $\alpha$  on  top of the doubly magic nuclei  $^{16}$O to describe $^{20}$Ne was considered 
using the generalized THSR wave function recently \cite{Bo,Bo2}. We will use the implementation of correlations according to the THSR approach 
\cite{THSR} that is able to unify clustering in nuclei with shell model approaches, if the parameters of the variational approach 
are chosen correspondingly. 
It is a challenge to present nuclear structure calculations to give a general in-medium description 
that contains both the limit of cluster formation at low densities, i.e. outside the nucleus, 
as well as the quasiparticle (shell-model) approach 
that is applicable at high densities, as already known from nuclear matter calculations in homogeneous systems \cite{SRS}, 
see also \cite{RSSN} for the four-nucleon case.

Shell model calculations tend to underestimate the decay width of $\alpha$ emitting nuclei like $^{212}$Po 
substantially \cite{Mang}. Preformation of $\alpha$-like correlations is indispensable \cite{Buck} 
to explain the observed decay widths. Cluster states have been considered already some time ago, see Ref. \cite{Arima}. 
Only recently systematic approaches have been considered that combine the shell model with cluster model calculations, 
see \cite{Varga,Delionbuch} and 
references given there. The preformation amplitude obtained there is in reasonable agreement with the experimental data, 
the amount of \{core + $\alpha$\} clustering amplitude in the parent state of about 30 \%
is found that is much higher than former microscopic estimates. 
A calculation using a modified Woods-Saxon potential has been published recently \cite{Delion}. 
In spite of the fact that the form of the single-particle potential is chosen ad hoc, the results are very reasonable. 
A microscopic approach leading to this empirical pocket-structure mean-field potential is, however, missing. 
Very recently \cite{Schuck13} it was shown that also in a restricted Hartree-Fock calculation
cluster formation can be described approximately, however, the separation of the c.o.m. motion 
has to be performed in a rigorous manner. For this, the single-particle approach must be improved treating
few-particle correlations responsible in forming bound states. 
%


After explaining the separation of the c.o.m.~motion in Section \ref{comSchr}, $\alpha$-like correlations are treated in Sec.~\ref{sec:3}.
Exploratory calculations for $^{212}$Po are presented in Sec.~\ref{expl} showing the formation of a pocket in the effective $\alpha$-cluster
potential near the surface of the double magic $^{208}$Pb core nucleus. Discussions and conclusions are drawn finally in Sec.~\ref{discussion}.

\section{The c.o.m. and intrinsic Schr\"odinger equations}
\label{comSchr}

We consider a few-body cluster, in particular $A_c$ nucleons of mass $m$ with two body interaction 
$V_{ij}({\bf r}_i,{\bf r}_j,{\bf r}'_i,{\bf r}'_j)$. Further details, such as isospin dependence of the interaction and of the masses, 
are neglected so that $m_n\approx m_p = m$. 
More details of the interaction potential will be discussed in the following sections.

To characterize the state of system, we can introduce the positions ${\bf r}_i$ (coordinate space representation)
or the momenta ${\bf p}_i$ (momentum space representation), whereas spin and isospin are not considered explicitly. 
If the interaction depends only on the relative positions ${\bf r}_i$ and there is no external potential, 
the problem is homogeneous in space and the total momentum is conserved. It is advantageous to introduce new observables,
the c.o.m. position ${\bf R}=\sum_i^{A_c}{\bf r}_i/A_c$, the relative coordinates ${\bf s}_j, \,\,j=1\dots A_c-1$, in particular 
Jacobian coordinates. Canonically conjugate momenta are the total momentum ${\bf P}= \sum_i^{A_c}{\bf p}_i$ and the relative momenta
${\bf k}_j, \,\,j=1\dots A_c-1$. 
As an example, for $A_c=4$ such transformations to Jacobi-Moshinsky coordinates are given in Sec. \ref{homogen}.

The introduction of the c.o.m. motion as a collective degree of freedom is also of general importance if we consider clusters
(bound states) consisting of $A_c$ particles. If the intrinsic interaction is strong compared with external influences from, e.g., 
core nuclei or homogeneous nuclear matter, 
such clusters can be considered as new elementary particles as it may happen at low density or when the cluster is quite far out 
in the surface of a nucleus. 
Then, the dynamical behavior is only given by the c.o.m. motion, whereas the intrinsic structure is nearly not changing.

In quantum theory, we try to subdivide the wave function $\Psi({\bf R},{\bf s}_j)$ into two parts,
\begin{equation}
\label{4}
\Psi({\bf R},{\bf s}_j)=\varphi^{\text{intr}}({\bf s}_j,{\bf R})\,\Phi({\bf R})
\end{equation}
This subdivision is unique (up to a phase factor $\Phi({\bf R})\to e^{i \alpha({\bf R})}\Phi({\bf R}),
 \varphi^{\text{intr}}({\bf s}_j,{\bf R}) \to e^{-i \alpha({\bf R})}  \varphi^{\text{intr}}({\bf s}_j,{\bf R})$) if,
 besides the normalization $\int dR\,ds_j\,|\Psi({\bf R},{\bf s}_j)|^2\equiv \int d^3R\,\int d^{3A-3}s_j\,|\Psi({\bf R},{\bf s}_j)|^2
 =1$, one also imposes the following individual normalisations (multiple integrals are not 
 indicated explicitly within the present Section)
\begin{equation}
\label{normS}
\int dR\,|\Phi({\bf R})|^2=1
\end{equation}
and for each ${\bf R}$
\begin{equation}
\label{normint}
\int ds_j |\varphi^{\text{intr}}({\bf s}_j,{\bf R})|^2=1\,.
\end{equation}

The Hamiltonian of a cluster 
may be written as 
\begin{equation}
H=\left(-\frac{\hbar^2}{2Am} \nabla_R^2+T[\nabla_{s_j}]\right)\delta({\bf R}-{\bf R}')\delta({\bf s}_j-{\bf s}'_j)
+V({\bf R},{\bf s}_j;{\bf R}',{\bf s}'_j)
\end{equation}
where the kinetic energy of the c.o.m. motion is explicitly given. The kinetic energy of the internal motion of the cluster, $T[\nabla_{s_j}]$,
depends on the choice of the Jacobi coordinates (see Sec. \ref{homogen} for $A=4$). 
The interaction $V({\bf R},{\bf s}_j;{\bf R}',{\bf s}'_j)$  contains the mutual interaction $V_{ij}({\bf r}_i,{\bf r}_j,{\bf r}'_i,{\bf r}'_j)$ 
between the particles 
as well as the 
interaction with an external potential (for instance, the potential of the core nucleus) 
and is, in general, non-local in space. We will specify the interaction $V({\bf R},{\bf s}_j;{\bf R}',{\bf s}'_j)$
when considering the $\alpha$ particle on top of a double magic core nucleus in Sec. \ref{sec:3} and Sec. \ref{expl}. At present, a local
external potential may be considered to explain the separation of the c.o.m. motion.

To find stationary states we take the expectation value of (4) with (1) and minimize
\begin{equation}
\label{5a}
\delta \left\{ \int dR\,ds_j\,dR'\,ds'_j\,\Psi^*({\bf R},{\bf s}_j) H \Psi({\bf R}',{\bf s}'_j) 
-E \int  dR |\Phi({\bf R})|^2-\int dR\,F({\bf R}) \int ds_j |\varphi^{\text{intr}}({\bf s}_j,{\bf R})|^2\right\}=0.
\end{equation}
The variation of the wave function is not restricted after the boundary conditions (\ref{normS}), (\ref{normint}) are
taken into account by the Lagrange parameters $E$ and $F({\bf R}) $.

The variation with respect to $\Phi^*({\bf R})$  yields the wave equation for the c.o.m. motion
\begin{eqnarray}
\label{9}
&&-\frac{\hbar^2}{2Am} \nabla_R^2\Phi({\bf R})-\frac{\hbar^2}{Am}\int ds_j \varphi^{\text{intr},*}({\bf s}_j,{\bf R}) 
[\nabla_R \varphi^{\text{intr}}({\bf s}_j,{\bf R})][\nabla_R\Phi({\bf R})]
\nonumber \\ &&
-\frac{\hbar^2}{2Am}\int ds_j \varphi^{\text{intr},*}({\bf s}_j,{\bf R}) 
[ \nabla_R^2 \varphi^{\text{intr}}({\bf s}_j,{\bf R})] \Phi({\bf R})
+\int dR'\,W({\bf R},{\bf R}')  \Phi({\bf R}')=E\,\Phi({\bf R})\,
\end{eqnarray}
with the c.o.m. potential
\begin{eqnarray}
\label{9c}
W({\bf R},{\bf R}')=\int ds_j\,ds'_j\,\varphi^{\text{intr},*}({\bf s}_j,{\bf R}) \left[T[\nabla_{s_j}]
\delta({\bf R}-{\bf R}')\delta({\bf s}_j-{\bf s}'_j)+V({\bf R},{\bf s}_j;{\bf R}',{\bf s}'_j)\right]
\varphi^{\text{intr}}({\bf s}'_j,{\bf R}')\,.
\end{eqnarray}

The variation of $\varphi^{\text{intr},*}({\bf s}_j,{\bf R})$ at fixed ${\bf R}$  yields the wave equation for the intrinsic motion
\begin{eqnarray}
\label{10}
&&-\frac{\hbar^2}{Am}  \Phi^*({\bf R}) [\nabla_R\Phi({\bf R})]
[\nabla_R \varphi^{\text{intr}}({\bf s}_j,{\bf R})]
-\frac{\hbar^2}{2Am}  |\Phi({\bf R})|^2
\nabla_R^2 \varphi^{\text{intr}}({\bf s}_j,{\bf R})
\nonumber \\ &&
+\int dR'\,ds'_j\, \Phi^*({\bf R}) \left[T[\nabla_{s_j}]
\delta({\bf R}-{\bf R}')\delta({\bf s}_j-{\bf s}'_j)+V({\bf R},{\bf s}_j;{\bf R}',{\bf s}'_j)\right]
 \Phi({\bf R}')\varphi^{\text{intr}}({\bf s}'_j,{\bf R}')=F({\bf R}) \varphi^{\text{intr}}({\bf s}_j,{\bf R})\,.
\end{eqnarray}
We emphasize that we should allow for non-local interactions. In particular, the Pauli blocking considered below
is non-local. Also the nucleon-nucleon interaction can be taken as non-local potential. 
To simplify the calculations, often local approximations are used for the potentials.
If in addition to the external potential also further conditions have to be implemented, further Lagrange multipliers are needed. 
For instance, the antisymmetrization with respect to the 
states of the core nucleus leads to a norm kernel $N$ \cite{Matsuse} to be considered in the following section \ref{sec:3}.

\section{The $\alpha$ particle case}
\label{sec:3}

\subsection{Quasi-particle representation}

We apply this formalism to the  $\alpha$ particle case, or more generally, to the correlation of four nucleons
moving in a nuclear system. The four nucleons are taken with different spin 
or isospin (not indicated explicitly in the following) that may form an $\alpha$ particle.  
The nucleon-nucleon interaction $ V_{N-N} $ will be specified below, see Eq. (\ref{sepa}).
Concerning the nuclear system we consider first nuclear matter (subsection \ref{homogen}). This case is comparatively
simple because it is homogeneous and the total momentum  ${\bf P}=\sum_i^4  {\bf p}_i$ of the few-particle system 
is conserved. After that we consider finite nuclei. For reasons to be discussed below,
the formation of an $\alpha$ particle on top of a double magic nucleus is of particular interest. 

In principle the theoretical formulation of an $\alpha$-cluster on top of a heavy doubly magic nucleus like $^{208}$Pb, 
the case to be considered in this work, is rather straight forward. In the so-called Tamm-Dancoff approximation (TDA), 
we consider the following Schr\"odinger equation
\begin{eqnarray}
\label{normkernel1}
&&(\varepsilon_{n_1} +\varepsilon_{n_2} +\varepsilon_{n_3} +\varepsilon_{n_4})\Psi^{\nu}_{n_1n_2n_3n_4} + 
\frac{1}{2}\sum_{n'_1n'_2}[1-f(\varepsilon_{n_1})] [1-f(\varepsilon_{n_2})]  \bar v_{n_1n_2n'_1n'_2}\Psi^{\nu}_{n'_1n'_2n_3n_4} 
\nonumber\\ &&+\,\, \mbox{permutations} = 
E_{\nu}N_{n_1n_2n_3n_4}\Psi^{\nu}_{n_1n_2n_3n_4}\,.
\end{eqnarray}
\noindent
The $\varepsilon_{n_i}$ are the single particle shell model energies corresponding to the mean field potential of the $^{208}$Pb core, 
that is $\hat h_i|n_i\rangle = \varepsilon_{n_i}|n_i\rangle$ where $\hat h$ is the single particle Hamiltonian of nucleons moving 
in the mean field of the lead core and $|n_i\rangle$ are the corresponding eigen functions. 
In this basis the antisymmetrised matrix elements of the two body force are given by $\bar v_{n_1n_2n_3n_4}$. 
Furthermore, 
the single-nucleon occupation ($\tau = n,p)$ is defined as 
\begin{equation}
\label{blocking1}
 f(\varepsilon_{n_\tau})= \Theta \left(\mu_{\tau}-\varepsilon_{n_\tau}\right),
\end{equation}
and
the projector on single particle states above the doubly magic core is given by
\begin{equation}
\label{normkernel2}
N_{n_1n_2n_3n_4} = \langle n_1n_2n_3n_4|\Theta(\hat h_1-\mu_1)\Theta(\hat h_2-\mu_2)
\Theta(\hat h_3-\mu_3)\Theta(\hat h_4-\mu_4)|n_1n_2n_3n_4\rangle
\end{equation}
\noindent
where $\Theta(x)$ is the step function and the $\mu_i$'s are the chemical potentials of the valence nucleons. 
Of course for the $\alpha$-like cluster considered here the chemical potentials are pairwise equal.

The above four particle Tamm-Dancoff equation is formally easy. However in the case of an $\alpha$ particle, 
i.e., an asymptotically  strongly bound cluster, the solution of this equation is absolutely non-trivial. 
The problem lies in the fact that one has to reproduce two limits correctly: on the one hand for 
the $\alpha$ particle being at large distances from the Pb core, 
the solution should contain the correct asymptotic limit of a lead core interacting only via the 
Coulomb force with an otherwise unperturbed $\alpha$. 
On the other hand, once the $\alpha$-like four nucleon cluster gets inside the Pb core, its cluster aspect gets 
dissolved and the four nucleons shall be described within the usual shell model approach. 
To have a consistent incorporation of both limits is, as well known, a very hard problem and has only 
been achieved so far within crude approximations \cite{Lovas,Delion}. 
A further very important aspect of the $\alpha$ particle cluster to be discussed in detail below in Sec. \ref{expl}, 
is the fact that in contrast to the case of the deuteron, 
the binding of the $\alpha$ particle gets lost quite abruptly once it enters the tail of the Pb core density. 
We have studied this effect in quite some detail in a series of earlier papers of $\alpha$ particles in low density 
nuclear matter \cite{RSSN,Sogo}. 
We think that the effect persists in finite systems. One could envisage to solve the above equation with a two center shell model, 
one for the $\alpha$ particle and the other for the lead core. However, this procedure also is not free of problems concerning 
for instance spurious center of mass motion, etc. In this work we will adopt a different strategy. 
Our focus will be how the $\alpha$ particle is modified entering from the outside into the region of finite density of the Pb core. 
We will treat the c.o.m. motion in Local Density Approximation (LDA). However, the intrinsic wave function 
of the $\alpha$ particle will be considered fully quantal. 
%
%

Within a quantum many-particle approach, the treatment of the interacting many-nucleon system needs some approximations 
that may be obtained in a consistent way from a Green functions approach.
In a first step, we can introduce the quasiparticle picture where the nucleons 
are moving independently in a mean field, described by a single-particle Hamiltonian $\hat h$ given above,
with single-nucleon states $|n_i\rangle$ as the shell states of the  $^{208}$Pb core.
In the next 
step we go beyond the quasi-particle picture and take the full interaction within the $A_c$-particle cluster into 
account. In the case of four nucleons considered here,
we have in position space representation 
\begin{eqnarray}
&&[E_4-\hat h_1 -\hat h_2- \hat h_3 - \hat h_4]\Psi_4({\bf r}_1 {\bf r}_2 {\bf r}_3{\bf r}_4)=
\int d^3 {\bf r}_1'\,d^3 {\bf r}_2' \langle {\bf r}_1{\bf r}_2|B \,\,V_{N-N}| {\bf r}_1'{\bf r}_2'\rangle
\Psi_4({\bf r}_1'{\bf r}_2'{\bf r}_3{\bf r}_4)\nonumber \\ && +
\int d^3 {\bf r}_1'\,\,d^3{\bf r}_3'  \langle {\bf r}_1{\bf r}_3|B \,\,V_{N-N}|
{\bf r}_1'{\bf r}_3'\rangle \Psi_4({\bf r}_1'{\bf r}_2{\bf r}_3'{\bf r}_4)
+ \text{four further permutations.}
\label{15}
\end{eqnarray}
The six  nucleon-nucleon interaction terms contain besides the nucleon-nucleon potential $V_{N-N}$ also the 
 blocking operator $B$ that can be given in quasi-particle state representation. 
For the first term on the r.h.s. of Eq. (\ref{15}), the expression
\begin{eqnarray}
 B(1,2)=[1-f_1(\hat h_1)-f_2(\hat h_2)]  
\label{15a}
\end{eqnarray}
results which is the typical blocking factor for the so-called particle-particle 
Random-Phase Approximation  (ppRPA) \cite{RS}.
The phase space occupation (we give the  internal quantum state $\nu=\sigma,\,\tau$ explicitly)
\begin{equation}
\label{occ}
f_\nu(\hat h) =\sum_n^{\text{occ.}}| n,\nu \rangle \langle n,\nu |
\end{equation}
indicates the phase space that according to the Pauli principle is not available for an interaction process of a nucleon 
with internal quantum state $\nu$.
 Here, we will use the Tamm-Dancoff (TDA) expression  $[1-f_1(\hat h_1)][1-f_2(\hat h_2)] $ 
 which neglects the hole-hole contributions that are of relevance in deriving the gap equation if pairing is considered. 
 In this way, our treatment is similar to the study of Cooper pairs by Cooper 
\cite{Cooper} that uses (\ref{15a}), only extended here to the case of quartets.
 We will not consider the Bogoliubov transformation introducing BCS quasiparticles 
 so that we discuss in the following the TDA expression. 
The Pauli blocking factor can be given in form of a projection operator 
 ${\cal P}^{\text{Pauli}}=1-\sum_n^{\text{occ.}}| n ,\nu\rangle \langle n,\nu |$ so that the quasiparticle 
subspace used to form the cluster is orthogonal to the subspace of the 
occupied shell model states in the core nucleus. Then, the norm kernel $N$ can be dropped.
 In homogeneous matter, the states below the Fermi energy are blocked out. In the Local Density Approximation (LDA) 
 used in the present work, the reduction of the phase space due to the Pauli principle is taken into account by the ansatz for
 the wave function, see Eq. (\ref{trial}). Note, however, that in the general case where the overlap between the 
occupied shell model states in the core nucleus and the wave function of the $\alpha$-like cluster remains finite,
the norm kernel (\ref{normkernel1}),  (\ref{normkernel2}) cannot be dropped. This problem shall be investigated in future work. 
 A systematic derivation of these expressions, also for the general case of finite temperatures, can be given 
using the Matsubara Green function method \cite{RSM,SRS}.

 Considering homogeneous nuclear matter characterized by the nucleon densities 
$n_\tau$ with $\tau = (n,p)$ (we drop the spin variable $\sigma$),
 the quasi-particle states are momentum eigenstates so that
the in-medium wave equation (\ref{15}) becomes simpler in momentum 
representation. The single-particle Hamiltonian  $\hat h_i$ as well as the Pauli blocking operator $B$ are diagonal in 
momentum representation, and Eq. (\ref{15}) reads  for the $\alpha$-like state (we will mark Fourier transformed 
quantities with a 'tilde')
\begin{eqnarray}
&&\left[\varepsilon_{\tau_1}^{\rm mf}({\bf p}_1)+ \varepsilon_{\tau_2}^{\rm mf}({\bf p}_2)
+\varepsilon_{\tau_3}^{\rm mf}({\bf p}_3)+\varepsilon_{\tau_4}^{\rm mf}({\bf p}_4)\right]
{\tilde \Psi}_4({\bf p}_1 {\bf p}_2{\bf p}_3{\bf p}_4)
\nonumber \\ &&+
\int \frac{d^3 {\bf p}_1'}{(2 \pi)^3}\int \frac{d^3 {\bf p}_2'}{(2 \pi)^3}\,
\left[1-f_{\tau_1}(\varepsilon_{\tau_1}^{\rm mf}({\bf p}_1))\right]\, 
\left[ 1-f_{\tau_2}(\varepsilon_{\tau_2}^{\rm mf}({\bf p}_2))\right]
\,{\tilde V}_{N-N}({\bf p}_1, {\bf p}_2;{\bf p}'_1,{\bf p}'_2)
{\tilde \Psi}_4({\bf p}'_1 {\bf p}'_2{\bf p}_3{\bf p}_4)
\nonumber \\ &&+ \text{five permutations} = E_4({\bf P}) 
{\tilde \Psi}_4({\bf p}_1 {\bf p}_2{\bf p}_3{\bf p}_4)\,.
\label{15aa}
\end{eqnarray}
Here, $\varepsilon_\tau^{\rm mf}({\bf p})=\hbar^2 p^2/2m +V_\tau^{\rm mf}({\bf p})$ 
contains the quasiparticle mean-field shift $V_\tau^{\rm mf}({\bf p})$,
and the Fermi function $f_\tau(E)=[\exp((E-\mu_\tau)/(k_BT)+1]^{-1}$ becomes the step function 
$\Theta(E_{\text{Fermi},\tau}-E)$ for zero temperature $T=0$, where $E_{\text{Fermi},\tau}=\mu_\tau$ 
denotes the Fermi energy of the neutrons or protons, see Eq. (\ref{blocking1}). Note that Eq. (\ref{15aa}) can be generalized for 
the case of finite temperatures $T$. Then, the energy eigenvalue $E_4$ as well as the wave function ${\tilde \Psi}_4$
will depend in addition to $n_n,\,n_p$ also on $T$.
The solution of this four-particle in-medium equation for homogeneous matter 
at arbitrary temperatures has been investigated extensively, see \cite{RSM,RSSN,Sogo,Roepke1,Ropke2011}.

We discuss the in-medium wave equation  (\ref{15aa}) more in detail. The medium modifications are originated by two effects:\\
i) The self-energy shifts $V_\tau^{\rm mf}({\bf p})$ contained in the single-particle Hamiltonian $\hat h_i$. We will denote
these contributions by the external part
\begin{equation}
\label{15aa0}
{\tilde V}^{(4),{\text{ext}}}({\bf p}_1 {\bf p}_2{\bf p}_3{\bf p}_4)=V_{\tau_1}^{\rm mf}({\bf p}_1)+
V_{\tau_2}^{\rm mf}({\bf p}_2)+V_{\tau_3}^{\rm mf}({\bf p}_3)+V_{\tau_4}^{\rm mf}({\bf p}_4)\,.
\end{equation}
ii) The Pauli blocking terms that modify the nucleon-nucleon interaction. We denote the interaction part including 
the Pauli blocking by the intrinsic part
\begin{eqnarray}
\label{15aaa}
{\tilde  V}^{(4),{\text{intr}}}({\bf p}_1 {\bf p}_2{\bf p}_3{\bf p}_4,{\bf p}_1' {\bf p}_2'{\bf p}_3'{\bf p}_4')&=&
\left[1-f_{\tau_1}(\varepsilon_{\tau_1}^{\rm mf}({\bf p}_1))\right]\, 
\left[ 1-f_{\tau_2}(\varepsilon_{\tau_2}^{\rm mf}({\bf p}_2))\right]
\,{\tilde V}_{N-N}({\bf p}_1, {\bf p}_2;{\bf p}'_1,{\bf p}'_2) \delta ({\bf p}_3-{\bf p}_3') \delta ({\bf p}_4-{\bf p}_4')
\nonumber \\ &&
\text{+ five permutations}\,,
\end{eqnarray}
the integrals in Eq. (\ref{15aa}) are modified correspondingly. The account of the Pauli blocking in the 
effective wave equation (\ref{15aa}) is indispensable to have a conserving approximation. Both,
the self-energy in mean-field approximation and the Pauli blocking given by the Fermi distribution 
are obtained in the approximation of an uncorrelated medium. Higher order approximations to the 
in-medium few-particle Green functions will improve the in-medium wave equation allowing for correlations in the medium as discussed in the 
last section \ref{discussion}.

\subsection{$\alpha$-like correlations in homogeneous nuclear matter}
\label{homogen}

To solve the four-nucleon problem separating the c.o.m. motion as a collective degree of freedom,
we introduce relative and c.o.m. Jacobi-Moshinsky coordinates  (for details see \cite{Roepke1})
 \begin{eqnarray}
  \label{Jac1}
&& {\bf r}_1={\bf R}+{\bf s}/2+{\bf s}_{12}/2, \qquad {\bf r}_2={\bf R}+{\bf s}/2-{\bf s}_{12}/2, \nonumber \\ 
&& {\bf r}_3={\bf R}-{\bf s}/2+{\bf s}_{34}/2, \qquad {\bf r}_4={\bf R}-{\bf s}/2-{\bf s}_{34}/2. 
\end{eqnarray}
In momentum space we have the conjugate Jacobi momenta
 \begin{eqnarray}
 \label{Jac2}
&& {\bf p}_1={\bf P}/4+{\bf k}/2+{\bf k}_{12}, \qquad {\bf p}_2={\bf P}/4+{\bf k}/2-{\bf k}_{12}, \nonumber \\ 
&& {\bf p}_3={\bf P}/4-{\bf k}/2+{\bf k}_{34}, \qquad {\bf p}_4={\bf P}/4-{\bf k}/2-{\bf k}_{34}. 
\end{eqnarray}

\subsubsection{Zero-density limit - the free $\alpha$ particle}
\label{freealpha}

To be more transparent, we consider first the free $\alpha$ particle, i.e. the zero density case $n_B=0$.
The ansatz (\ref{4}) reads now (we denote the free case by the index 0)
\begin{equation}
\Psi_0({\bf R},{\bf s},{\bf s}_{12},{\bf s}_{34})=\varphi_0^{\text{intr}}({\bf s},{\bf s}_{12},{\bf s}_{34})\Phi_0({\bf R})\,.
\end{equation}
The Hamiltonian in position representation contains the intrinsic kinetic energy 
\begin{equation}
\label{kin4}
T_4[\nabla_{s_j}]=-\frac{\hbar^2}{2 m} \frac{\partial^2}{\partial {\bf s}^2}-\frac{\hbar^2}{ m} \frac{\partial^2}{\partial {\bf s}_{12}^2}
-\frac{\hbar^2}{ m} \frac{\partial^2}{\partial {\bf s}_{34}^2}
\end{equation}
and the nucleon-nucleon interaction potential $V^{(4),{\text{intr}}}({\bf s},{\bf s}_{12},{\bf s}_{34};{\bf s}',{\bf s}'_{12},{\bf s}'_{34})$ 
depending on intrinsic 
coordinates only. The potential $V^{(4),{\text{intr}}}$ contains six pair interaction terms, see Eq. (\ref{15}), 
where the Cartesian coordinates 
are transformed to Jacobian coordinates according to (\ref{Jac1}). In the homogeneous system, there is no external force acting on the 
$\alpha$ particle. 
Mean-field self-energy shifts and Pauli blocking vanish in the zero density limit.

Since the interaction does not contain any dependence on $ {\bf R} $, the intrinsic wave function
$\varphi^{\rm intr}_0( {\bf s}, {\bf s}_{12}, {\bf s}_{34})$ is also not depending on $ {\bf R} $ 
(a trivial phase factor $e^{i \alpha( {\bf R} )}$ can be eliminated, as discussed above, below Eq.~(\ref{4})).
The system of wave equations (\ref{9}), (\ref{10}) is considerably simplified. With respect to the application to homogeneous 
matter it is convenient to use the momentum representation. For the free $\alpha$ particle (zero-density limit),
Eq. (\ref{10}) reads
\begin{eqnarray}
\label{10a}
 &&\frac{\hbar^2}{2m} [k^2+2 k_{12}^2+2 k_{34}^2 ]\tilde \varphi_0^{\text{intr}}({\bf k},{\bf k}_{12},{\bf k}_{34})+
 \int \frac{d^3 k'}{(2 \pi)^3}\,\frac{d^3 k'_{12}}{(2 \pi)^3}\,\frac{d^3 k'_{34}}{(2 \pi)^3} {\tilde V}^{(4),{\text{intr}}} 
 ({\bf k},{\bf k}_{12},{\bf k}_{34};{\bf k}',{\bf k}'_{12},{\bf k}'_{34}) 
 \tilde \varphi_0^{\text{intr}}({\bf k}',{\bf k}'_{12},{\bf k}'_{34})
 \nonumber \\ &&
 = E_\alpha^{(0)} \tilde \varphi_0^{\text{intr}}({\bf k},{\bf k}_{12},{\bf k}_{34})\,.
 \end{eqnarray}
 Here, the four-nucleon interaction $V^{(4),{\text{intr}}}$ contains the six pair interactions in the free $\alpha$ cluster.
 The new Lagrange parameter $E_\alpha^{(0)}=F({\bf R})/|\Phi({\bf R})|^2$ coincides with the intrinsic energy of the
 free $\alpha$ particle (of course, independent of $\bf R$),
\begin{eqnarray}
\label{10bb}
&&E_\alpha^{(0)} =\int \frac{d^3 k}{(2 \pi)^3}\,\frac{d^3 k_{12}}{(2 \pi)^3}\,\frac{d^3 k_{34}}{(2 \pi)^3}\frac{\hbar^2}{2m} 
[k^2+2 k_{12}^2+2 k_{34}^2 ]
|\tilde \varphi_0^{\text{intr}}({\bf k},{\bf k}_{12},{\bf k}_{34})|^2
 \\ &&
+\int \frac{d^3 k}{(2 \pi)^3}\,\frac{d^3 k_{12}}{(2 \pi)^3}\,\frac{d^3 k_{34}}{(2 \pi)^3} \,
\frac{d^3 k'}{(2 \pi)^3}\,\frac{d^3 k'_{12}}{(2 \pi)^3}\,\frac{d^3 k'_{34}}{(2 \pi)^3}
\tilde \varphi_0^{\text{intr,*}}({\bf k},{\bf k}_{12},{\bf k}_{34})
V^{(4),{\text{intr}}}({\bf k},{\bf k}_{12},{\bf k}_{34};{\bf k}',{\bf k}'_{12},{\bf k}'_{34}) 
\tilde \varphi_0^{\text{intr}}({\bf k}',{\bf k}'_{12},{\bf k}'_{34})\,.\nonumber
\end{eqnarray}
The empirical value is $E_\alpha^{(0)}=-B_\alpha=-28.3$ MeV. 

In the free  $\alpha$ particle case, the c.o.m. potential $W({\bf R},{\bf R}')$, Eq. (\ref{9c}), is local. 
According to Eq. (\ref{10bb}), it reads $W({\bf R},{\bf R}')= E_\alpha^{(0)} \delta({\bf R}-{\bf R}') $ 
so that, in the zero density case,
Eq. (\ref{9}) reads
\begin{eqnarray}
\label{9a}
\frac{\hbar^2 P^2}{8 m} \tilde\Phi_0({\bf P})+E_\alpha^{(0)}\tilde\Phi_0({\bf P})=E_0({\bf P})\, \tilde\Phi_0({\bf P})\,.
\end{eqnarray}

 Equation (\ref{10a}) is the Schr\"odinger equation for the intrinsic motion of the $\alpha$ particle, and
 Eq. (\ref{9a}) is the  Schr\"odinger equation for the c.o.m. motion. The Lagrange parameter 
 $E_0({\bf P})\equiv \hbar^2/(8m) P^2+E_\alpha^{(0)}$ has the meaning of the total energy of the $\alpha$ particle. 
 
The wave functions $\Phi_0({\bf R}), \varphi^{\rm intr}_0( {\bf s}, {\bf s}_{12}, {\bf s}_{34})$ or their Fourier transforms follow
solving the Schr\"odinger equations. The solution for the c.o.m. motion, Eq. (\ref{9a}), is trivial in the case of homogeneous matter. 
In position representation 
results a plain wave with wave vector ${\bf P}$. To solve the wave equation for the intrinsic motion (\ref{10a})
we have to define the interaction. We choose a separable interaction \cite{RSSN}
with Gaussian form factor.
\begin{equation}
\label{sepa}
V_{N-N}({\bf p}_1, {\bf p}_2;{\bf p}'_1,{\bf p}'_2)=\lambda {\rm e}^{-({\bf p}_1-{\bf p}_2)^2/4 \gamma^2} {\rm e}^{-({\bf p}'_1-{\bf p}'_2)^2/4 
\gamma^2}\delta({\bf p}_1+{\bf p}_2-{\bf p}'_1-{\bf p}'_2)
\end{equation}
and find the approximate solution from a variational approach.

In particular, for the  Gaussian wave functions as a simple variational ansatz, the c.o.m. motion can be easily separated.  
For vanishing c.o.m. motion,
${\bf P} = 0$, we have for the internal wave function
\begin{equation}
{\tilde \varphi}^{\rm intr}_0 ({\bf p}_1,{\bf p}_2,{\bf p}_3, {\bf p}_4) = \frac{1}{\rm norm}
{\rm e}^{-(p^2_1+ p^2_2+ p^3_3+p^2_4)/b^2} \delta({\bf p}_1+{\bf p}_2+{\bf p}_3+{\bf p}_4),
\end{equation}
with the normalization $\sum_{p_1,p_2, p_3, p_4} |{\tilde \varphi}^{\rm intr}_0({\bf p}_1,{\bf p}_2,{\bf p}_3, {\bf p}_4)|^2 = 1$,
or, explicitly,
\begin{equation}
\label{Gauss}
\tilde \varphi^{\rm intr}_0({\bf k},{\bf k}_{12},{\bf k}_{34})=\frac{2^{6} (2 \pi)^{9/4}}{b^{9/2}}  {\rm e}^{-2{\bf k}^2_{12}/b^2}  {\rm e}^{-2{\bf k}^2_{34}/b^2}  
{\rm e}^{-{\bf k}^2/b^2}.
\end{equation}
With potential parameters $\lambda = -1449.6$ MeV  fm$^{3}$ and $\gamma = 1.152$ fm$^{-1}$ in Eq. (\ref{sepa}), 
the binding energy and rms radius of the 
free $\alpha$ particle  are reproduced, using the Gaussian variational ansatz for the intrinsic motion.
To show this, we calculate the intrinsic energy according to Eq. (\ref{10bb})
\begin{equation}
{\hat E}_\alpha^{(0)}(b)= \frac{9}{8}\frac{\hbar^2}{m} b^2 + 6 \lambda \frac{\gamma^6 b^3}{\pi^{3/2}(b^2+2 \gamma^2)^3}
\end{equation}
with a minimum $E^{(0)}_{\alpha} = -28.3$ MeV for the ground state energy at $b=1.034$ fm$^{-1}$.
The parameter $b$ reproduces the nucleonic point rms radius $\sqrt{\langle r^2 \rangle}=1.45$ fm
as $b^2=9/(4 \langle r^2 \rangle )= 1.069$ fm$^{-2}$.

\subsubsection{$\alpha$-like correlations in homogeneous nuclear matter at finite densities}

We continue to discuss the case of homogeneous nuclear matter that is of relevance when later introducing a local density approach.
In homogeneous systems, the separation of the c.o.m. motion is exact because the c.o.m. momentum is conserved. 
Since there the effective c.o.m. potential 
$W({\bf R},{\bf R}')$ depends only on ${\bf R}-{\bf R}'$ and gradient terms like $\nabla_R \,\varphi^{\rm intr}$ can be dropped,
we have from Eq. (\ref{9})
\begin{eqnarray}
\label{9aa}
&&-\frac{\hbar^2}{8m} \nabla_R^2\Phi({\bf R})
+\int d^3R'\,W({\bf R}-{\bf R}')  \Phi({\bf R}')=E_4\,\Phi({\bf R})\,.
\end{eqnarray}
after Fourier transform (remember, the transformed quantities are marked with a 'tilde')
\begin{eqnarray}
&&\left[-\frac{\hbar^2}{8 m} {\bf P}^2+ {\tilde W}({\bf P}) \right]
{\tilde \Phi}({\bf P})=E_4({\bf P})\,{\tilde \Phi}({\bf P})\,.
\label{21a}
\end{eqnarray}
To identify different contributions to the effective c.o.m. potential ${\tilde W}({\bf P})$ we consider the in-medium wave equation 
(\ref{15aa}) given in momentum representation. The eigenvalue $E_4({\bf P})$ will depend on the total momentum ${\bf P}$ not only 
due to the kinetic energies of the single-nucleon states, $ P^2/(8m)$ but also due to the mean-field shifts $V_\tau^{\rm mf}({\bf p})$
that may be put into the chemical potential in the rigid shift approximation, as well as due to the Pauli blocking terms $f_\tau(\varepsilon_\tau^{\rm mf}({\bf p}))$.
(Approximately, this dependence of  $E_4({\bf P})$ on $P$ can be described introducing an effective mass of the $\alpha$-like cluster.)

The wave equation for the intrinsic motion (\ref{10}) becomes also simplified in homogeneous systems
\begin{eqnarray}
\label{10aa}
T_4[\nabla_{s_j}] \varphi^{\text{intr}}({\bf s}_j,{\bf R})
+\int d^3R'\,d^9s'_j V^{(4)}({\bf R},{\bf s}_j;{\bf R}',{\bf s}'_j) \frac{\Phi({\bf R}')}{\Phi({\bf R})} \varphi^{\text{intr}}({\bf s}'_j,{\bf R}')
=\frac{F({\bf R})}{|\Phi({\bf R})|^2} \varphi^{\text{intr}}({\bf s}_j,{\bf R})\,.
\end{eqnarray}
Whereas the intrinsic kinetic energy $T_4[\nabla_{s_j}] $ is given by (\ref{kin4}), the interaction $V^{(4)}({\bf R},{\bf s}_j;{\bf R}',{\bf s}'_j)$ contains 
in addition to the mutual interaction also the medium effects, in particular the self-energy shifts and the Pauli blocking terms. 
For homogeneous systems it is convenient to pass over to momentum representation. With the Jacobi-Moshinsky momenta (\ref{Jac2}),
Eq. (\ref{15aa}) reads now (cf. Eqs. (\ref{10}), (\ref{10a}))
\begin{eqnarray}
\label{10ab}
 &&\frac{\hbar^2}{2m} [k^2+2 k_{12}^2+2 k_{34}^2 ]\tilde \varphi^{\text{intr}}({\bf k},{\bf k}_{12},{\bf k}_{34},{\bf P})+
 \int \frac{d^3 k'}{(2 \pi)^3}\,\frac{d^3 k'_{12}}{(2 \pi)^3}\,\frac{d^3 k'_{34}}{(2 \pi)^3} {\tilde V}^{(4)} 
 ({\bf k},{\bf k}_{12},{\bf k}_{34};{\bf k}',{\bf k}'_{12},{\bf k}'_{34};{\bf P}) 
 \tilde \varphi^{\text{intr}}({\bf k}',{\bf k}'_{12},{\bf k}'_{34},{\bf P})
 \nonumber \\ &&
 = {\tilde W}({\bf P}) \tilde \varphi^{\text{intr}}({\bf k},{\bf k}_{12},{\bf k}_{34},{\bf P})\,.
 \end{eqnarray}
 We used that for homogeneous systems the interaction term $ {\tilde V}^{(4)}$ is diagonal  with respect to the total momentum ${\bf P}$.
 The new Lagrange parameter ${\tilde W}({\bf P})$ is the Fourier transform of $F({\bf R})/|\Phi({\bf R})|^2$ and can be considered 
 as the intrinsic energy of the four-nucleon system.

The effective in-medium interaction ${\tilde V}^{(4)} ({\bf k},{\bf k}_{12},{\bf k}_{34};{\bf k}',{\bf k}'_{12},{\bf k}'_{34};{\bf P}) $ contains 
the external part (\ref{15aa0}) as well as the intrinsic part  (\ref{15aaa})  (to be transformed to Jacobi-Moshinsky momenta).
In addition to the terms 
that describe the intrinsic motion of the free $\alpha$ particle, additional contributions arise from the single-nucleon self-energy shift
$V_\tau^{\rm mf}$ and the Pauli blocking term $f_\tau[\varepsilon_\tau^{\rm mf}({\bf p})]$.
Accordingly, we decompose the effective c.o.m. potential 
\begin{equation}
{\tilde W}({\bf P})={\tilde W}^{\text{ext}}({\bf P})+{\tilde W}^{\text{intr}}({\bf P})
\end{equation}
into an external part ${\tilde W}^{\text{ext}}({\bf P})$, collecting the mean 
field shifts $V_\tau^{\rm mf}$ of the surrounding matter, and an intrinsic part 
${\tilde W}^{\text{intr}}({\bf P})$ that contains the intrinsic kinetic energy as 
well as the mutual interaction of the constituents including the Pauli blocking.
As seen from Eqs. (\ref{15aa}), (\ref{15aa0}),  the quasiparticle mean-field shift $V^{\rm mf}_\tau({\bf p})$ gives the first contribution
\begin{eqnarray}
\label{W4}
&&{\tilde W}^{\rm ext}({\bf P})=\int \frac{d^3 k}{(2 \pi)^3}\,\frac{d^3 k_{12}}
{(2 \pi)^3}\,\frac{d^3 k_{34}}{(2 \pi)^3}
 |{\tilde \varphi}^{\rm intr}({\bf k},{\bf k}_{12},{\bf k}_{34},{\bf P})|^2
 \nonumber \\ && \times\left[V^{\rm mf}_{\tau_1}\left(\frac{{\bf P}}{4}+\frac{
{\bf k}}{2}+{\bf k}_{12}\right)
 +V^{\rm mf}_{\tau_1}\left(\frac{{\bf P}}{4}+\frac{{\bf k}}{2}-{\bf k}_{12}\right)
 +V^{\rm mf}_{\tau_1}\left(\frac{{\bf P}}{4}-\frac{{\bf k}}{2}+{\bf k}_{34}\right)
 +V^{\rm mf}_{\tau_1}\left(\frac{{\bf P}}{4}-\frac{{\bf k}}{2}-{\bf k}_{34}
\right)\right]
\end{eqnarray}
to the Fourier transform of the four-particle c.o.m. potential $W({\bf R},
{\bf R}')$, Eq. (7), that depends for homogeneous systems only
on ${\bf R}-{\bf R}'$. This term acts on the free nucleons in quasiparticle 
states as well as on the bound nucleons in the cluster. If the momentum dependence
of the mean-field shift $V^{\rm mf}_{\tau}({\bf p})$ can be neglected (rigid 
shift approximation), both the scattering states as well as the bound 
four-nucleon states are shifted by the same amount. Then, the contribution to 
the shift of the binding energy (the difference between scattering
state and bound state energies) is cancelled.  
Simple 
approximations for the mean-field shifts in homogeneous matter are, e.g., 
given by Skyrme forces or relativistic mean-field energy shifts and are not 
discussed here in detail. 
Let us, however, mention that the mean field shifts are most of the time 
incorporated into a rigid shift not depending on ${\bf p}$ 
and an effective mass that give only a small contribution,  see \cite{Roepke1} 
for further details.
For finite nuclei, expressions for the mean-field shift like the Woods-Saxon 
potential are given in Sec. \ref{expl}.

A second contribution to the influence of the surrounding matter on the four-
nucleon system in Eq. (\ref{15aa}) is due to Pauli blocking,
given by the occupation $f_\tau[\varepsilon_\tau^{\rm mf}({\bf p})]$ of single 
quasiparticle nucleon states. As already given above, Eq. (\ref{blocking1}) and below  Eq. (\ref{15aa}),
in homogeneous matter (no dependence on ${\bf R}$), we 
adopt the single-nucleon occupation ($\tau = n,p)$ as 
\begin{equation}
\label{blocking}
 f_{\tau,{\bf p}}= f_\tau[\varepsilon_\tau^{\rm mf}({\bf p})]=f({\bf p};\mu_{\tau},T=0)=\Theta 
\left(\mu_{\tau}-\varepsilon_\tau^{\rm mf}({\bf p})\right).
\end{equation}
The chemical potentials $\mu_\tau$ coincide at zero temperature with the Fermi 
energy, 
$\mu_\tau=E_{\text{Fermi},\tau}= (\hbar^2/2 m) (3 \pi^2 n_\tau)^{2/3}$, and are 
determined by the respective densities.

The evaluation of the Pauli blocking term for arbitrary temperatures and 
arbitrary c.o.m. momenta ${\bf P}$ 
has been given in Ref. \cite{Ropke2011}.
Some special results for the zero temperature case that are not discussed 
in  \cite{Ropke2011} are given below.
For the Pauli blocking, we will consider the wave equation (\ref{15aa}) for 
zero total momentum, 
${\bf p}_1+{\bf p}_2+{\bf p}_3+{\bf p}_4=0$.
Note again that we replaced the RPA blocking term $[1-f_{\tau_1}
(\varepsilon_{p_1})-f_{\tau_2}(\varepsilon_{p_2})]$ 
by the TDA term $[1-f_{\tau_1}(\varepsilon_{p_1})][1-f_{\tau_2}(\varepsilon_{p_2})]$ 
which excludes the participation of already occupied 
single-particle states (below the Fermi surface) from the propagation of the 
four-nucleon state. 
The medium is treated as uncorrelated, 
and also the formation of a BCS state is excluded.

\subsubsection{Energy of intrinsic motion in homogeneous matter at  $ P=0$}
\label{intrinsichom}

We can expand ${\tilde W}({\bf P})$ with respect to ${\bf P}$ but in this work we only evaluate the  terms for ${\bf P}=0$. 
For the external part 
${\tilde W}^{\text{ext}}({\bf P})$ the higher orders in  ${\bf P}$ are zero if the mean-field potential is local. In general, as well known, 
within a gradient expansion the next term can be absorbed introducing effective masses. In particular, the mean-field shift 
$V^{\rm mf}_{\tau}({\bf p})$ in a homogeneous system can be treated this way leading to a rigid shift $V^{\rm mf}_{\tau}(0)$
and to the introduction of an effective nucleon mass $m^*$. We discuss here only the lowest order of the expansion with respect to the
single nucleon momentum ${\bf p}$. The introduction of the effective nucleon mass is straightforward, see \cite{Roepke1} where corresponding 
expressions for the homogeneous case are given.

The in-medium wave equation (\ref{15aa}) can be given in a Hermitean form and can be solved with a variational approach. 
After a projected product ansatz, self-consistent equations to solve the single-nucleon wave function are considered in 
Ref. \cite{Sogo}. For simplicity, here we use a Gaussian ansatz, see Eq. (\ref{Gauss}), that reads
\begin{equation}
\label{Gauss1}
\tilde \varphi^{\rm intr}({\bf p}_1,{\bf p}_2,{\bf p}_3, {\bf p}_4)=\frac{1}{\rm norm}  
\varphi_{\tau_1}({\bf p}_{1})\varphi_{\tau_1}({\bf p}_{2})\varphi_{\tau_1}({\bf p}_{3})\varphi_{\tau_1}({\bf p}_{4}) 
\delta({\bf p}_1+{\bf p}_2+{\bf p}_3+{\bf p}_4)
\end{equation}
with
\begin{equation}
\label{trial}
 \varphi_{\tau}({\bf p})={\rm e}^{-{\bf p}^2/b^2} \Theta\left[p-p_{\rm Fermi, \tau}\right]
\end{equation}
so that the Fermi sphere $p_{\rm Fermi, \tau}=(3 \pi^2 n_\tau)^{1/3}$ is blocked out, $b$ is a variational parameter.
To simplify the calculations we average the Fermi energies with respect to the isospin $\tau =n,p$ (symmetric matter), 
so that we perform the calculations for an excluded Fermi sphere 
$p_{\rm Fermi}=(3 \pi^2 n_B/2)^{1/3}$ with the total baryon density $n_B=n_n+n_p$.

Within the variational calculation, we have to evaluate the norm of the trial function (\ref{trial}) as well as the 
kinetic and potential energy. The Pauli blocking is already taken into account by the choice of the trial wave function
and must not be considered anymore. After transforming to the internal 
Jacobian coordinates ${\bf k},{\bf k}_{12},{\bf k}_{34}$, one has to perform multiple integrals, see App. \ref{app}.

The intrinsic motion of the four-nucleon system contains the kinetic energy 
and the interaction energy within the cluster taking into account Pauli blocking. 
Besides the shift ${\tilde W}^{\text{ext}}$ that acts on the nucleons both in the scattering (single nucleon-) states as well as 
in bound states, the dependence of effective c.o.m. 
potential (\ref{9c}) 
$ {\tilde W}={\tilde W}^{\text{ext}}+ {\tilde W}^{\text{intr}} $ on the c.o.m. momentum ${\bf P}$ and the baryon density
is determined by the internal part ${\tilde W}^{\text{intr}}$ that is sensitive to the formation of bound states.. 
The dependence of ${\tilde W}^{\text{intr}}({\bf P})$ on  ${\bf P}$ is due to the Pauli blocking term $B$ and has been considered 
in detail in \cite{Ropke2011}. Here we restrict us to the value ${\tilde W}^{\text{intr}}$ at  ${\bf P}=0$. Using Eq. (\ref{10ab}), we 
separate the mean-field shifts from ${\tilde V}^{(4)}  ({\bf k},{\bf k}_{12},{\bf k}_{34};{\bf k}',{\bf k}'_{12},
{\bf k}'_{34};{\bf P}) $ that give the contribution ${\tilde W}^{\text{ext}}$.
The in-medium equation for the intrinsic part of the $\alpha$ particle wave function is given by
\begin{eqnarray}
&& \left({\tilde W}^{\text{intr}}-\frac{\hbar^2}{2m} \left[ k^2+2 k_{12}^2+2 k_{34}^2 \right] \right)
{\tilde \varphi}^{\text{intr}}_4({\bf k}, {\bf k}_{12}, {\bf k}_{34})
\nonumber \\&&
=\int\frac{d^3 k'}{(2 \pi)^3}\frac{d^3 k_{12}'}{(2 \pi)^3}\frac{d^3 k_{34}'}{(2 \pi)^3} 
V_4^{\rm intr}({\bf k}, {\bf k}_{12}, {\bf k}_{34},{\bf k}', 
{\bf k}'_{12}, {\bf k}'_{34},{\bf P}=0){\tilde \varphi}^{\text{intr}}_4({\bf k}', {\bf k}'_{12}, {\bf k}'_{34}) \,,
\label{21b} 
\end{eqnarray}
where the four-nucleon interaction term $V_4^{\rm intr}$ contains also the Pauli blocking terms for ${\bf P}=0$, 
see Eqs. (\ref{15}), (\ref{15a}). The explicit form is obtained from 
\begin{equation}
\label{28aa}
V_4^{\rm intr}({\bf p}_1,{\bf p}_2,{\bf p}_3,{\bf p}_4,{\bf p}'_1,{\bf p}'_2,
{\bf p}'_3,{\bf p}'_4) 
=[1-f({\bf p}_1)][1-f({\bf p}_2)] V_{N-N}({\bf p}_1,{\bf p}_2;{\bf p}'_1,
{\bf p}'_2) \delta({\bf p}_3-{\bf p}'_3)\delta({\bf p}_4-{\bf p}'_4)
+ {\text{five permutations}}
 \end{equation}
 after transforming to Jacobian momenta (\ref{Jac2}). 
 A solution of this equation within a variational approach is described for the free $\alpha$ particle in the previous subsection  \ref{freealpha}. 
We will do the same at finite density with the variational ansatz (\ref{Gauss1}),  (\ref{trial}), see also App. \ref{app}.
In contrast to the expression (\ref{10bb}) for the zero-density case, for 
arbitrary ${\bf P}$ the minimum of
\begin{eqnarray}
\label{varhom}
&&{\tilde W}({\bf P})=\frac{\hbar^2}{2m} \int \frac{d^3 k}{(2 \pi)^3}\,
\frac{d^3 k_{12}}{(2 \pi)^3}\,\frac{d^3 k_{34}}{(2 \pi)^3}
\left[ k^2+2 k_{12}^2+2 k_{34}^2 \right]|{\tilde \varphi}^{\rm intr}({\bf k},
{\bf k}_{12},{\bf k}_{34},{\bf P})|^2 
+ {\tilde W}^{\text{ext}}({\bf P})
 \\ &&
+\int \frac{d^3 k}{(2 \pi)^3}\,\frac{d^3 k_{12}}{(2 \pi)^3}\,\frac{d^3 k_{34}}
{(2 \pi)^3}
\,\frac{d^3 k'}{(2 \pi)^3}\,\frac{d^3 k'_{12}}{(2 \pi)^3}\,\frac{d^3 k'_{34}}
{(2 \pi)^3}
{\tilde \varphi}^{\rm intr,*}({\bf k},{\bf k}_{12},{\bf k}_{34},{\bf P}) 
V_4^{\rm intr}({\bf k}, {\bf k}_{12}, {\bf k}_{34},{\bf k}', 
{\bf k}'_{12}, {\bf k}'_{34},{\bf P})
{\tilde \varphi}^{\rm intr}({\bf k}',{\bf k}'_{12},{\bf k}'_{34},{\bf P})\nonumber
\end{eqnarray}
has to be found  with Eqs. (\ref{Gauss1}), (\ref{trial}) and $b$ again the single variational 
parameter. ${\tilde W}^{\text{ext}}({\bf P})$ is given by (\ref{W4}) and $V_4^{\rm intr}$ is given by (\ref{28aa}) with arbitrary ${\bf P}$.

\begin{figure}[th] 
 	\includegraphics[width=0.7\textwidth]{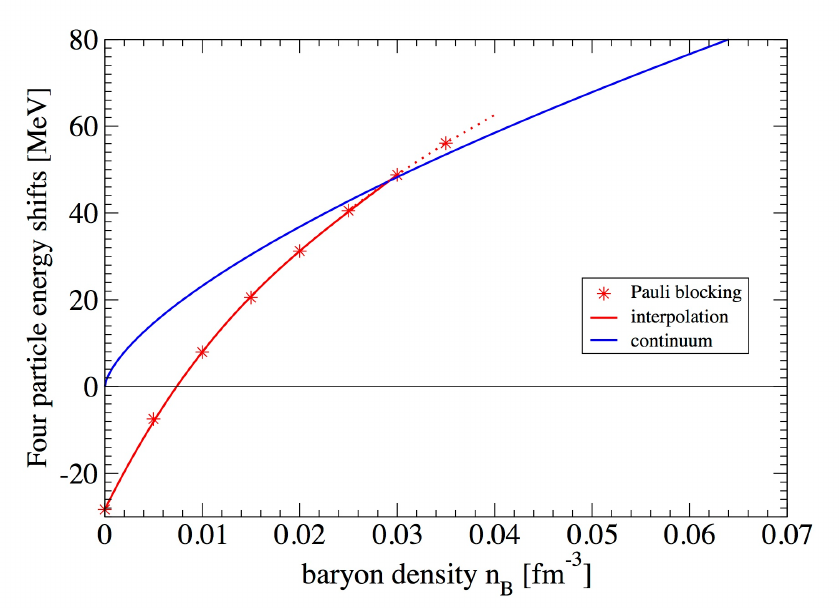}
	\caption{Internal four-nucleon energy (no c.o.m. motion) in a medium with nucleon density $n_B=n_n+n_p$. 
	The continuum edge of free single-particle states is given by $4 E_{\rm Fermi}$, Eq. (\ref{4Fermi}). 
	At zero baryon density, the four-nucleon energy is given by the binding energy of the $\alpha$ particle, 
	$E^0_\alpha = -B^0_\alpha =-28.3$ MeV. With increasing density, the binding energy $B^0_\alpha$ is 
	decreasing due to the Pauli blocking, Eq. (\ref{bound}) (stars). The four-nucleon bound state disappears at
	$n_B \approx 0.03$ fm$^{-3}$. A fit to the calculated values, Eq. (\ref{Paulifit}), is also shown}
 \label{Fig:4nuc} 
 \end{figure}

 In contrast to the free $\alpha$ particle case detailed in Sec. \ref{freealpha},
the  eigenvalue ${\tilde W}^{\text{intr}}$ now becomes dependent on the density that enters the Pauli blocking via the Fermi momentum.
The results are shown in Fig. \ref{Fig:4nuc}. An interpolation formula that reproduces these results is given below in  Eq. (\ref{Paulifit}).

We discuss the result for ${\tilde W}^{\text{intr}}$ in more detail. To add four nucleons (neutrons and protons, two spin orientations) to
nuclear matter with density $n_\tau$ we consider two cases which are based on the scenario which we described in the introduction, 
that is the $\alpha$ particle as a bound state only exists in the far surface. 
As soon as the $\alpha$ enters the region of higher density, its binding fades away and the four nucleons go over into shell model states 
(eventually with pairing). Therefore: \\
i) At first, the four nucleons are treated as free, uncorrelated particles that corresponds to the shell model states. The minimum energy (the 
edge of the continuum of scattering states) necessary to introduce the nucleons is 
\begin{equation}
\label{4Fermi}
 {\tilde W}^{\rm intr, free}[n_\tau]=2 E_{\text{Fermi}}(n_n)+2 E_{\text{Fermi}}(n_p)
 =\frac{\hbar^2}{ m}\left[(3 \pi^2 n_n)^{2/3}+(3 \pi^2 n_n)^{2/3}\right]\,.
\end{equation}
The four nucleons are introduced at the corresponding Fermi energy with zero total momentum. Only the kinetic energy is 
needed to determine the edge of the continuum of scattering states.
This four-particle energy for single-nucleon states is shown for symmetric matter ($n_n=n_p=n_B/2$) 
in Fig. \ref{Fig:4nuc} with the (blue) line starting at zero energy.\\
ii) Below the continuum of scattering states, bound states may occur in the four-nucleon system at very low densities. 
In the zero density limit,we have the formation of the $\alpha$ particle at the bound state energy 
$E_\alpha^{(0)} = -28.3$ MeV for the internal motion, the energy of
the c.o.m. motion vanishes at ${\bf P}=0$. The energy of the four-nucleon bound state is shifted at finite density of the surrounding 
nuclear matter due to Pauli blocking so that 
\begin{equation}
\label{bound}
 {\tilde W}^{\rm intr, bound}[n_\tau]=E_\alpha^{(0)}+{\tilde W}^{\text{Pauli}}(n_\tau)
 =-28.3\,\, {\text{MeV}}+{\tilde W}^{\text{Pauli}}(n_\tau)\,.
\end{equation}
The Pauli blocking shift ${\tilde W}^{\text{Pauli}}(n_\tau)$ for nuclear matter 
is caused by the terms containing the phase space occupations
$f_\tau(E)$.

The minimum of the energy leads, with increasing density, 
to a wave function $ {\tilde \varphi}_{\tau}({\bf p})$  that has near the Fermi momentum a sharp maximum for the 
distribution of the occupation of the single-nucleon states. It is expected for any added nucleons that, at 
minimum energy, it occupies the Fermi momentum if the interaction is neglected. We obtain a 
solution at the continuum edge of single-particle states at high densities, whereas below a critical value 
$n_{B,\text{cluster}} \approx 0.03$ fm$^{-3}$ a bound state is formed.
The corresponding energies as a function of density are shown in Fig. \ref{Fig:4nuc} with the red asterisk's. 
Note that the sharp appearance of a bound state at a critical density where blue and red lines cross is possibly 
a consequence of the simple variational ansatz that contains only one parameter $b$. Until now, there is no exact solution of the 
four-particle problem near the so-called Mott point \cite{RSM} where, 
due to Pauli blocking, the bound state is dissolved in the continuum of scattering states. The same applies also for finite temperatures 
discussing, for instance, the disappearance of quartetting with increasing density \cite{RSSN} that 
seems to be a sharp transition to pairing. In principle one cannot exclude, however, a fast but smooth merging of both 
solutions.

As just discussed, in contrast to the two-nucleon case where the pairing solution exists also 
in the degenerate case, the $\alpha$-like four-nucleon bound state 
may disappear abruptly at $n_{B,\text{cluster}}$ what can be explained considering the density of states near the Fermi energy \cite{RSSN}. 
Let us discuss this difference in more detail. Supposing that 
the c.o.m. of the 
particles is at rest (${\bf P}=0$), we obtain for the two particle case
\begin{equation}
g_{2}(\omega = 2\mu) \propto \int d^3P \int d^3 k \bar n_{{\bf P}/2 - {\bf k}}\bar 
n_k \delta(2 \mu -e_{{\bf P}/2 - {\bf k}} - e_k) \delta({\bf P}) = 
\propto \sqrt{\mu}
\end{equation}
where $\bar n_k = 1-n_k$ with $n_k = \Theta(\mu - e_k)$ and $e_k = 
\frac{k^2}{2m}$.

Analogously we obtain for the four particle level density at the Fermi 
energy with total c.o.m at rest
\begin{eqnarray}
&&g_4(\omega = 4\mu) \propto \int d^3 P d^3 P' d^3 k  d^3 k'\bar n_{{\bf P}/2 - 
{\bf k}}
\bar n_{{\bf P}2 + {\bf k}}\bar n_{{\bf P}'2 - {\bf k}'}
\bar n_{{\bf P}'/2 + {\bf k}'}
\nonumber \\ && \times 
\delta (4 \mu -e_{{\bf P}/2 - {\bf k}}-e_{{\bf P}/2 + {\bf k}}-e_{{\bf P}'/2 - {\bf k}'}-
e_{{\bf P}'/2 + {\bf k}'})
\delta({\bf P}+{\bf P}') =0\,.
\end{eqnarray}
We see that in the four particle case, for positive $\mu$, energy conservation and phase space constraint cannot be 
fulfilled simultaneously and, thus, no four particle correlations can build up around the Fermi energy. 
This is a quite dramatic difference to the two particle case where the level density remains finite at the Fermi level. 
For negative $\mu$, i.e., for the case where there is binding, the Fermi step $n_k$ is zero and no qualitative difference 
between two and four particle cases exists. The two particle case is, therefore, very exceptional with respect to all heavier clusters. 
Therefore, when the $\alpha$ particle approaches the $^{208}$Pb core,
the internal structure of the $\alpha$-like cluster remains relatively stable until it is dissolved quite abruptly at the 
critical density $n_{B,\text{cluster}}=0.03$ fm$^{-3}$ which is very low. In addition to the deformation by the Fermi momentum, described above, that no states in 
momentum space are occupied below the Fermi level, the change in the variational 
parameter $b$ that describes the width of the Gaussian wave function, is moderate. It changes 
from its value $b = 1.034$ fm$^{-1}$ at $n_B=0$ to $b = 0.84$ fm$^{-1}$ at $n_{B,{\rm cluster}}=0.03$ fm$^{-3}$.

In conclusion, considering homogeneous nuclear matter, additional nucleons (two neutrons, two protons) can form an $\alpha$-like cluster. 
In the zero density limit  the binding energy amounts 28.3 MeV.
As soon as the density takes a finite value, due to the Pauli blocking the binding energy is shifted.
Bound states are possible for $n_B \le 0.03$ fm$^{-3}$. To give a simple relation for the dependence on the baryon density,
the fit formula derived  within a variational approach to solve the in-medium four-nucleon wave equation,
\begin{equation}
\label{Paulifit}
{\tilde W}^{\rm Pauli}(n_B) =4515.9\, n_B -100935\, n_B^2+1202538\, n_B^3
\end{equation}
can be used, $n_n=n_p=n_B/2$. For  $n_B \ge 0.03$ fm$^{-3}$, no bound state is 
formed, and the four nucleons added to the lead core nucleus  
are implemented on top of the Fermi energy $\mu$, see Fig. \ref{Fig:4nuc}.

The intrinsic wave function (\ref{Gauss1}), (\ref{trial}) is ${\bf R}$ dependent via the Fermi momentum 
if the inhomogeneous case is considered, 
for instance an $\alpha$ particle on top of a heavy nucleus whose c.o.m. position is fixed at ${\bf R}_{\text{core}}=0$. 
Also the intrinsic energy $W({\bf R})$ introduced in Eq. (\ref{9}) becomes depending on ${\bf R}$
via  $n_\tau({\bf R})$. This will be discussed in Sec. \ref{expl} with the introduction of an effective potential 
for the $\alpha$-like state near the lead core in $^{212}$Po.
 
 \subsection{$\alpha$-like correlations in a nucleus, Thomas-Fermi approximation}
 \label{TFnucl}

 Now we discuss the formation of $\alpha$-like correlations for a finite 
nuclear system, in particular the nucleus $^{212}$Po considered below. 
Now, a mean-field potential $V^{\rm mf}_\tau({\bf r})$ acts on the nucleons, taken as 
local and depending on isospin $\tau$. 
As well-known from the shell model, a harmonic oscillator potential or a 
Woods-Saxon
like potential can be used to determine single-nucleon orbits that are 
occupied up to the Fermi energy. Often 
this potential is considered as a local one, only depending on the nucleon 
coordinate ${\bf r}$. 
For comparison, in the homogeneous case considered before, any dependence on  ${\bf r}$ 
disappears, and the mean-field contribution is
a constant that can be added to the intrinsic energy.

The solution of  the four-nucleon system using the c.o.m. coordinate $\bf R$ as new degree of freedom
as well as relative coordinates is not as simple as in the homogeneous case.
We start from the general expressions given in Sec. \ref{comSchr}. 
In particular, we neglect the terms containing $\nabla_R \varphi_4^{\text{intr}}
({\bf s}_j,{\bf R})$ so that Eqs. (\ref{9}), (\ref{10})
reduce to
 \begin{eqnarray}
\label{9b}
&&-\frac{\hbar^2}{8m} \nabla_R^2\Phi({\bf R})
+\int d^3R'\,W({\bf R},{\bf R}') \,\, \Phi({\bf R}')=E_4\,\Phi({\bf R})\,
\end{eqnarray}
with the effective c.o.m. potential
 \begin{eqnarray}
\label{9bb}
&&W({\bf R},{\bf R}')=\int d^9s_j\,d^9s'_j\,\varphi_4^{\text{intr},*}({\bf s}_j,{\bf R}) \left[T_4[\nabla_{s_j}]
\delta({\bf R}-{\bf R}')\delta({\bf s}_j-{\bf s}'_j)+V_4({\bf R},{\bf s}_j;{\bf R}',{\bf s}'_j)\right]
\varphi_4^{\text{intr}}({\bf s}'_j,{\bf R}')\,.
\end{eqnarray}
The in-medium four-particle interaction $V_4({\bf R},{\bf s}_j;{\bf R}',
{\bf s}'_j) $ follows from Eq. (\ref{15aa}).
Besides the intrinsic nucleon-nucleon interaction $V_{N-N}$ it contains also two medium effects, 
the quasiparticle mean-field shift $V_\tau^{\rm mf}({\bf r})$ that leads to the contribution ${W}^{\rm ext}({\bf R, R'})$,
see Eqs. (\ref{W4}) and (\ref{11b}) below, and the Pauli blocking terms 
$\propto f_{\tau}(\varepsilon_\tau^{\rm mf})V_{N-N}$ that leads to the contribution ${W}^{\rm Pauli}({\bf R, R'})$, see Eqs. 
(\ref{21b}), (\ref{bound}). Both the contributions ${W}^{\rm ext}({\bf R, R'})$, ${W}^{\rm Pauli}({\bf R, R'})$ depend on the density 
of the nuclear medium and vanish for the free $\alpha$ particle case. In general, these medium contributions are non-local 
and  depend on ${\bf R, R'}$.

The variation of the functional (\ref{5a}) with respect to 
$\varphi_4^{\text{intr},*}({\bf s}_j,{\bf R})$ at fixed ${\bf R}$ yields
\begin{eqnarray}
\label{10b}
&&
\int d^3R'\,d^9s'_j\,  \left[T_4[\nabla_{s_j}] \delta({\bf R}-{\bf R}')\delta({\bf s}_j-{\bf s}'_j)+V_4({\bf R},
{\bf s}_j;{\bf R}',{\bf s}'_j)\right]
 \frac{\Phi({\bf R}')}{|\Phi({\bf R})|^2}\varphi_4^{\text{intr}}({\bf s}'_j,{\bf R}')
 =E_4^{\rm intr}({\bf R}) \varphi_4^{\text{intr}}({\bf s}_j,{\bf R})\,
 \end{eqnarray}
where we introduced the intrinsic energy $E_4^{\rm intr}({\bf R})=F({\bf R})/|\Phi({\bf R})|^2$ in analogy to 
Eqs. (\ref{10bb}), (\ref{10ab}). In contrast to the free $\alpha$-particle energy $E_\alpha^{(0)}$, 
the intrinsic energy contains in-medium effects
and depends on the c.o.m. position ${\bf R}$. If the effective c.o.m. 
potential $W({\bf R},{\bf R}')$ is taken in local approximation, we have $W({\bf R},{\bf R}')=E_4^{\rm intr}({\bf R}) 
\delta({\bf R}-{\bf R}')$.   

In general, these equations are non-local in ${\bf R}$ space due to the 
potential energy $V_4({\bf R},{\bf s}_j;{\bf R}',{\bf s}'_j)$
that contains the mean-field contribution $V^{\rm ext}_4$  defined below 
as well as the intrinsic interaction $V^{\rm intr}_4$ within the four-nucleon cluster 
(c.f. also Eqs. (\ref{15aa0}) and (\ref{15aaa})),
\begin{equation}
\label{W4R}
 V_4({\bf R},{\bf s}_j;{\bf R}',{\bf s}'_j)=V^{\rm ext}_4({\bf R},{\bf s}_j;
{\bf R}',{\bf s}'_j)
 +V^{\rm intr}_4({\bf R},{\bf s}_j;{\bf R}',{\bf s}'_j)\,.
\end{equation}
We discuss both contributions separately  together with some approximations.

Usually, the mean field of the nucleus is taken as local in position space, 
neglecting momentum dependence what makes also ${W}^{\rm ext}({\bf R, R'})$ local. 
Below we use the Woods-Saxon potential $V^{\rm mf}_\tau ({\bf r})$
that depends on the position ${\bf r}_i$ of the four nucleons, $\tau =n,p$. 
Transforming to Jacobi coordinates we have for the interaction with an external (mean-field) potential 
 \begin{eqnarray}
\label{11b}
&& V^{\rm ext}_4({\bf R},{\bf s}_j;{\bf R}',{\bf s}'_j)=\left[V_{\tau_1}^{\rm mf}({\bf R}+\frac{1}{2}{\bf s}+\frac{1}{2}{\bf s}_{12})
+V_{\tau_2}^{\rm mf}({\bf R}+\frac{1}{2}{\bf s}-\frac{1}{2}{\bf s}_{12})
\right. \nonumber \\ && \left.
+V_{\tau_3}^{\rm mf}({\bf R}-\frac{1}{2}{\bf s}+\frac{1}{2}{\bf s}_{34})+V_{\tau_4}^{\rm mf}({\bf R}-\frac{1}{2}{\bf s}-\frac{1}{2}{\bf s}_{34})
 \right] \delta({\bf R}-{\bf R}') \delta({\bf s}-{\bf s}')  \delta({\bf s}_{12}-{\bf s}'_{12}) \delta({\bf s}_{34}-{\bf s}'_{34})\,.
 \end{eqnarray}
 For the effective c.o.m. potential $W({\bf R},{\bf R}')=W^{\rm ext}({\bf R}) \delta({\bf R}-{\bf R}')+W^{\rm intr}({\bf R},{\bf R}')$
 we have the mean-field contribution
 \begin{equation}
 \label{wext}
W^{\rm ext}({\bf R})=\int d^3s\, d^3s_{12}\, d^3s_{34}\,|\varphi_4^{\text{intr}}({\bf s}_j,{\bf R})|^2 
V^{\rm ext}_4({\bf R},{\bf s}_j;{\bf R},{\bf s}_j)\,.
\end{equation}
Similar to the introduction of a double-folding potential, the effective c.o.m. interaction term due to the mean-field potential 
follows after averaging with the intrinsic density distribution.

As a further component to the effective c.o.m. potential energy, the Pauli blocking appears. 
The Pauli principle as consequence of 
antisymmetrization means that states below the Fermi energy 
are blocked if further nucleons are added to the lead core (we consider $^{212}$Po). We denote this as the intrinsic four-particle energy
$ {\tilde W}^{\rm intr}[n_\tau({\bf R})]$, Eqs. (\ref{21b}), (\ref{28aa}), 
(and more explicitly Eqs. (\ref{bound}), (\ref{Paulifit})) of the intrinsic motion that is a functional of the nucleon density 
$ n_\tau({\bf R})$ 
of the surrounding medium. We will obtain this local density expressions 
within a more general approach that is able to 
go also beyond the local density approximation (LDA) and makes the terms that 
are neglected in the LDA more transparent. In principle, the full quantal solution may be possible.
Here we will consider 
the Thomas-Fermi approximation as a simple approximation to the LDA.
The reader not interested in the technical details of how to get to 
LDA and the approximations involved, can directly jump to Eq. (\ref{10c}) where the 
same expression for the intrinsic energy as in (\ref{varhom}) is given, only in LDA. 

The intrinsic interaction including blocking terms $B$ reads in position space ($i=1\dots 4$) (cf. Eq. (\ref{28aa}) in momentum representation)
 \begin{eqnarray}
\label{11c}
 V^{\rm intr}_4({\bf r}_i;{\bf r}'_i)=&&\int d^3r''_1\,d^3r_2'' \,\,
\langle {\bf r}_1 {\bf r}_2 |[1-f_1(\varepsilon_{n_1})][1-f_2(\varepsilon_{n_2})]|{\bf r}''_1 {\bf r}''_2 \rangle
\langle {\bf r}''_1 {\bf r}''_2 |V_{N-N}|{\bf r}'_1 {\bf r}'_2 \rangle \delta({\bf r}'_3-{\bf r}_3) \delta({\bf r}'_4-{\bf r}_4)
\nonumber \\ &&
+ {\text{ five permutations}}
 \end{eqnarray}
where $\langle {\bf r}_1|f_1(\varepsilon_{n_1})|{\bf r}'\rangle$ that is defined with the single-nucleon quasiparticle states
$\psi_n({\bf r})$, is given in a local approximation in the following.

We can introduce Jacobi coordinates to separate the c.o.m. motion and perform a Fourier transform to momentum representation.
As above, the nucleon-nucleon interaction can be taken in a separable form so that
\begin{eqnarray}
 \langle {\bf r}''_1 {\bf r}''_2 |V_{N-N}|{\bf r}'_1 {\bf r}'_2 \rangle \delta(
{\bf r}'_3-{\bf r}''_3) \delta({\bf r}'_4-{\bf r}''_4)
&& =\int \frac{d^3 k'_{12}}{(2 \pi)^3}\, \frac{d^3 k''_{12}}{(2 \pi)^3} 
{\rm e}^{i {\bf s}''_{12} \cdot {\bf k}''_{12} - i {\bf s}'_{12} \cdot {\bf k}'_{12}} 
 V_{N-N}({\bf k}''_{12};{\bf k}'_{12}) \delta({\bf s}'-{\bf s}'') \delta(
{\bf s}'_{34}-{\bf s}''_{34})
 \nonumber \\ && 
 =\langle {\bf s}'', {\bf s}''_{12}, {\bf s}''_{34} |V_{N-N}|{\bf s}', 
{\bf s}'_{12}, {\bf s}'_{34} \rangle\,.
\end{eqnarray}

More difficult is the treatment of the Pauli blocking term $B$ that is an exchange term and non-local in position space. 
We delegate it to App. \ref{app:2} where the corresponding approximations are given. This allows us in future work to 
eliminate some of the approximation made here.

We recover in Thomas-Fermi approximation the expression for the shift given in the homogeneous case,
only with the parametric dependence on the c.o.m. position ${\bf R}$ via the baryon density $n_B({\bf R})$. 
Though we gave here the whole series of approximations leading in the end to LDA or TF expressions, 
where, in principle corrections can be evaluated, we will give below general arguments in favor of such a 
local procedure for the c.o.m. motion of the $\alpha$ particle.

After the local approximation with respect to  ${\bf R}$ was introduced, we solve Eq. (\ref{10b}) 
within a variational approach. 
With Eq. (\ref{wext}) that contains also the intrinsic wave function, the minimum of the functional
\begin{eqnarray}
\label{10cc}
 &&
\left[W^{\rm ext}_4({\bf R})+\int d^9s_j\,\varphi_4^{\text{intr},*}({\bf s}_j,
{\bf R})T_4[\nabla_{s_j}]
\varphi_4^{\text{intr}}({\bf s}_j,{\bf R})
\right.  \\ && \left.
+\int d^9s_j  d^9s'_j d^9s''_j\,\varphi_4^{\text{intr},*}({\bf s}_j,{\bf R}) 
B({\bf R},{\bf s}_j,{\bf s}'_j)\,V^{(4)}_{N-N}({\bf s}'_j,{\bf s}''_j) 
\varphi_4^{\text{intr}}({\bf s}''_j,{\bf R})\right]\left[\int d^9s_j\,|
\varphi_4^{\text{intr}}({\bf s}_j,{\bf R})|^2\right]^{-1}
=E^{\rm intr}_4({\bf R})\nonumber
\end{eqnarray}
within a given set of functions $\varphi_4^{\text{intr}}({\bf s}_j,{\bf R})$ gives an approximation for the intrinsic wave function and 
the intrinsic energy. The Pauli blocking term $B$ depends on the position ${\bf R}$. In the approximation considered here, it is diagonal
in momentum representation, and the dependence on ${\bf s}_j,{\bf s}'_j$ follows after Fourier transform as shown in Eq. (\ref{F1}).

In the following Section we perform exploratory calculations with the separable interaction given above. 
It is of advantage to use a mixed representation where the intrinsic part is given in momentum representation.
Again we use the Fermi blocked Gaussian ansatz (\ref{trial}) for the intrinsic wave function with the width parameter 
as the only variational input (which becomes density, and, via the local density, also {\bf R} dependent).
The intrinsic interaction and the Pauli blocking gives for the contribution to the potential due to the interaction between the 
nucleons 1 and 2 (the other five follow from permutations and gives rise to the factor six below). More explicitly, Eq. (\ref{10cc}) reads
\begin{eqnarray}
\label{10c}
&& \left[ W^{\rm ext}({\bf R})
+\frac{\hbar^2}{2m}  \int \frac{d^3k}{(2 \pi)^3}  \frac{d^3k_{12}}{(2 \pi)^3} \frac{d^3k_{34}}
{(2 \pi)^3}\left[ k^2+2 k_{12}^2+2 k_{34}^2 \right]  \,
 |{\tilde \varphi}_4^{\text{intr}}({\bf k},{\bf k}_{12},{\bf k}_{34},{\bf R})|^2
 \right. \nonumber \\  && \left.
 +6 \int \frac{d^3k}{(2 \pi)^3}  \frac{d^3k_{12}}{(2 \pi)^3}   \frac{d^3k'_{12}}{(2 \pi)^3}  \frac{d^3k_{34}}{(2 \pi)^3} \,
{\tilde \varphi}_4^{\text{intr},*}({\bf k},{\bf k}_{12},{\bf k}_{34},{\bf R}) \left[1-f_1({\bf R},\frac{\bf k}{2}+{\bf k}_{12})\right] 
\left[1-f_2({\bf R},
\frac{\bf k}{2}-{\bf k}_{12})\right]\,
\right.\nonumber \\ && \left.
 \times V_{N-N}({\bf k}_{12},{\bf k}'_{12}) {\tilde \varphi}_4^{\text{intr}}({\bf k},{\bf k}'_{12},{\bf k}_{34},{\bf R})\right]
\left[\int \frac{d^3k}{(2 \pi)^3}  \frac{d^3k_{12}}{(2 \pi)^3} \frac{d^3k_{34}}{(2 \pi)^3}
\,|{\tilde \varphi}_4^{\text{intr}}({\bf k},{\bf k}_{12},{\bf k}_{34},{\bf R})|^2
\right]^{-1}=
E^{\rm intr}_4({\bf R})\,.
\end{eqnarray}
This is now the LDA version of (\ref{varhom}). We emphasize that the approach  given here allows to improve the LDA. 
In particular, the Pauli blocking is only approximately determined by the baryon density $n_B({\bf R})$ at the c.o.m. position ${\bf R}$.
As in discussed in context with Eq. (\ref{RR"}), the baryon density and, correspondingly, the Fermi momentum $p_{\rm Fermi}(n_B)$
has to be averaged over the neighborhood of ${\bf R}$ corresponding to the spatial extension of the intrinsic wave function 
$ \varphi_4^{\text{intr}}({\bf s},{\bf s}_{12},{\bf s}_{34},{\bf R})$. 
An improvement of the LDA is, e.g., given when for $f_1^{\text{Wigner}}$ the actual position 
such as ${\bf R}+({\bf s}+{\bf s}_{12})/2$ is taken.

The interaction term contains the Pauli blocking that is simple in momentum 
representation since it is diagonal in that representation.
For ${\bf P}=0$ we used within a variational approach a Gaussian internal 
wave function 
${\tilde \varphi}^{\text{intr}}_4({\bf k}, {\bf k}_{12}, {\bf k}_{34};{\bf R})$  
where in the phase space
$\{ {\bf k}, {\bf k}_{12}, {\bf k}_{34}\}$ the volume $|{\bf k}+{\bf k}_{12}/2|
\le k_{\text{Fermi}}({\bf R}),\,\,
|{\bf k}-{\bf k}_{12}/2|\le k_{\text{Fermi}}({\bf R}),\,\,|{\bf k}+{\bf k}_{34}/2|
\le k_{\text{Fermi}}({\bf R}),\,\, {\text{and}} \,
|{\bf k}-{\bf k}_{34}/2|\le k_{\text{Fermi}}({\bf R})$ is excluded. Consequently, 
the variational ansatz for the
internal wave function should vanish within that excluded volume. 
The blocking term is taken in local density (Fermi gas) approximation. 
The nucleon-nucleon interaction $V_{N-N}$ without blocking terms gives the 
bound state energy of the $\alpha$ 
particle $E^{(0)}_\alpha = - 28.3$ MeV. 
We separate this part so that $W^{\rm int}_4({\bf R})=E^{(0)}_\alpha+W^{\text{Pauli}}(
{\bf R})$. 
The dependence of $W^{\text{Pauli}}({\bf R})$ on the surrounding baryon density  
$n_B({\bf R})$ is given by Eq. (\ref{Paulifit}). 
This nucleon density $n_B({\bf R})$ is determined by the core nucleus that may be 
described in shell model calculation. 

Note that the Thomas-Fermi approximation given here, {\it i.e.} the 
introduction of a 'local momentum',
 is possible because the inverse width parameter  $b$ of the intrinsic wave 
function 
of the $\alpha$-like bound state remains nearly unchanged, it is reduced only 
by 17 \% when it merges with the quasi-continuum of 
shell-model single quasi-particle states. This means that the $\alpha$ 
particle, even up to the break up point, remains a 
rather compact entity with small extension, of the same order as the surface 
width of the core nucleus entailing 
that a local approach can be used at least as a first reasonable attempt. 
This is quite opposite to the pairing case 
where the size of the Cooper pairs can be as large as the nucleus itself 
invalidating a LDA approach.  
The derivation given here allows to go beyond LDA if the corresponding 
approximations are improved. In principle also a fully quantal solution can be envisaged.

\section{Exploratory calculations}
\label{expl}

\begin{figure}[th] 
 	\includegraphics[width=0.5\textwidth]{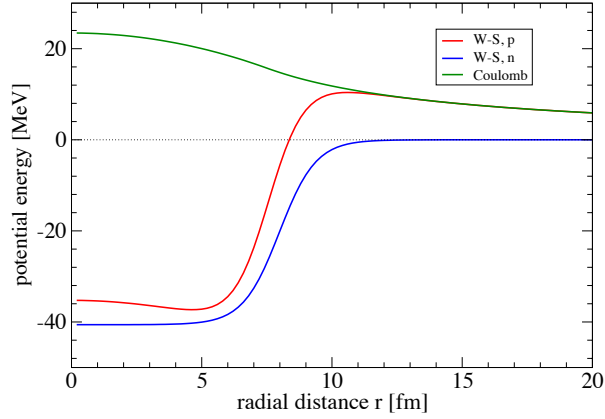}
	\caption{Coulomb potential and isospin dependent Woods-Saxon potentials for the  $^{208}$Pb core.}
 \label{Fig:2nuca} 
 \end{figure}

For demonstration we consider $^{212}$Po, i.e. an $\alpha$ particle on top of the double magic $^{208}$Pb core nucleus \cite{Delion1}.
We take Woods-Saxon mean field potentials \cite{Delion,Dudek,Rost} that are used for the description of nuclei in the lead region. 
In particular, for the neutrons of the $^{208}$Pb core we use
\begin{equation}
\label{WSn}
V^{\rm mf}_n( r )=-\frac{40.6}{1+{\rm e}^{(r-R_n)/a}}
\end{equation}
with $R_n=1.347 A^{1/3}=7.891$ fm  and $a = 0.7$ fm. 
For the protons we take
\begin{equation}
\label{WSp}
V^{\rm mf}_p( r )=-\frac{58.7}{1+{\rm e}^{(r-R_p)/a}}+V^{\rm Coul}( r),
\end{equation}
with $R_p=1.275 A^{1/3}=7.554$ fm and $a = 0.7$ fm. 
The Coulomb potential produced by the lead core is taken for a homogeneously charged sphere as (units in MeV, fm)
\begin{equation}
\label{Coul}
V^{\rm Coul}( r)=82 \frac{1.44}{r},\,\,\,r>R_p;\qquad 
V^{\rm Coul}( r)=82 \frac{1.44}{R_p}\left[ \frac{3}{2}- \frac{1}{2} \frac{r^2}{R_p^2}\right],\,\,\,r<R_p.
\end{equation}
These potentials are shown in Fig. \ref{Fig:2nuca}, see also Fig. 2 of Ref. \cite{Delion}.

On the two-neutron, two-proton cluster ($\alpha$-like cluster) acts the potential given by Eqs. (\ref{wext}), (\ref{11b}). 
As a local approximation we take the mean-field potential at the c.o.m. position ${\bf R}$, 
i.e. $2 V^{\rm mf}_n( R )+2V^{\rm mf}_p( R )+2V^{\rm Coul}( R )$ to simplify the calculations, 
but avoiding to perform the spatial average with the intrinsic wave function. The deviation
$2 V^{\rm mf}_n( R )+2V^{\rm mf}_p( R )+2V^{\rm Coul}( R )$ and a correction $\Delta V^{\rm ext}( R )$ according to
\begin{equation}
\label{WSa}
\Delta V^{\rm ext}({\bf R} )=W^{\rm ext}({\bf R})-[2 V^{\rm mf}_n({\bf R} )+2V^{\rm mf}_p({\bf R} )+2V^{\rm Coul}({\bf R} )] 
\end{equation}
is of interest in the low-density region $n_B \le 0.03$ fm$^{-3}$ 
where $\alpha$-like bound states can be formed, but it is assumed to be small because the potentials are smooth and the $\alpha$ particle is
well localized in co-ordinate space so that this correction $\Delta V^{\rm ext}( R )$ can be neglected.
The local approximation where the mean-field potential $W^{\rm ext}({\bf R} )$ is replaced by the sum of 
the mean-field potentials of the four constituents at the c.o.m. position ${\bf R}$ can be improved taking into account 
the correction $\Delta V^{\rm ext}({\bf R} )$.

For the internal part $W_4^{\rm intr}( R )$ of the c.o.m. potential we have to estimate the baryon density $n_B( R)$  
that is responsible for the Pauli blocking. To be consistent within the local density approach given here, we use the Thomas-Fermi 
approximation in the average baryon potential $W^{\rm ext}_4( R )/4$,
\begin{equation}
\label{nbar}
n_B( R )=\frac{2}{3 \pi^2}\left[\frac{2 m}{\hbar^2} \left(\mu-\frac{1}{4}W^{\rm ext}_{4}( R )\right)\right]^{3/2}.
\end{equation}
From the mass number $A=\int n_B( R ) d^3R =208$ of the core nucleus, 
the value $\mu=-5.504$ MeV is obtained for the chemical potential (Fermi energy).

We consider the case of inhomogeneous nuclear matter where, compared with the homogeneous case, the c.o.m. motion is not trivial. 
Instead of Eq. (\ref{21a}) for the homogeneous case we have now  from Eq. (\ref{9})
\begin{eqnarray}
&&\left[-\frac{\hbar^2}{8 m} \frac{\partial^2}{\partial {\bf R}^2}+W({\bf R}) \right]
\Phi({\bf R})
=E_4 \Phi({\bf R})\,
\label{21c}
\end{eqnarray}
with
\begin{equation}
\label{Vcm}
W({\bf R})=E^{\text{intr}}_4({\bf R})={ W}^{\rm ext}({\bf R})+W^{\text{intr}}({\bf R})={ W}^{\rm ext}({\bf R})
+E^{(0)}_\alpha+W^{\rm Pauli}(\bf R)\,.
\end{equation}
Note that in general the effective c.o.m. potential $W({\bf R})$
is not local in space but depends on two variables ${\bf R}$ and ${\bf R'}$. 

\begin{figure}[hb] 
	\includegraphics[width=0.8\textwidth]{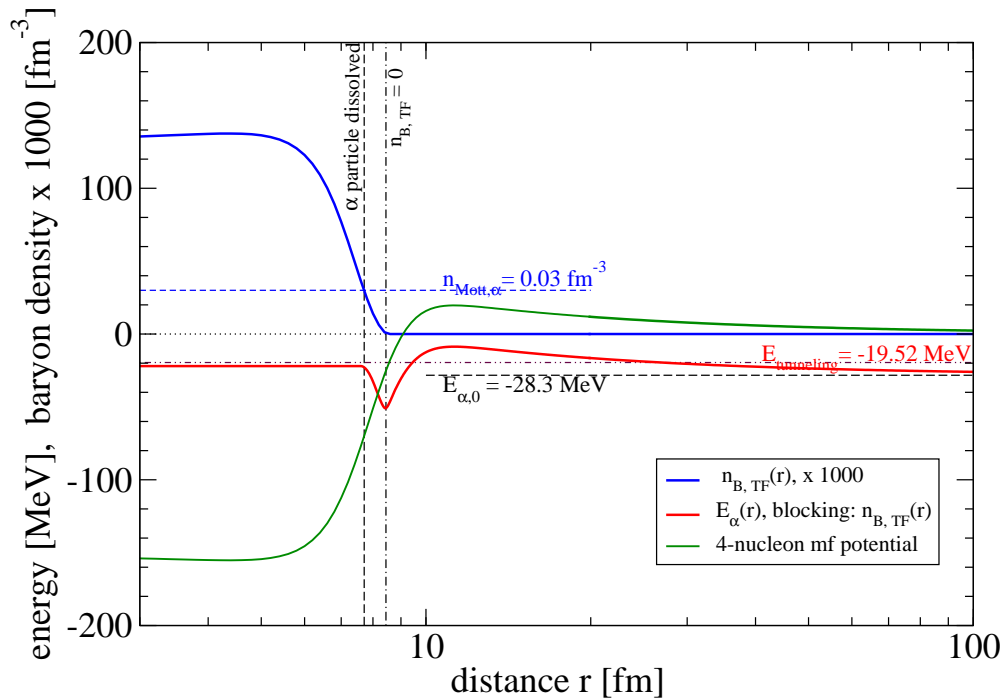}
	\caption{Local effective potential  $W({\bf R})$ (\ref{Vcm}) (red, full) )
	with respect to the lead $^{208}$Pb core
	for the Woods-Saxon 
	like distribution (\ref{WSa}), $A=208$. The Thomas-Fermi density and the Fermi energy $4 E_{\rm Fermi}$ of the four
	added nucleons is shown, 
	furthermore the measured energy ($Q$-value) of the emitted $\alpha$ particle. 
	The distance at which the density becomes the critical value $n_B=0.0292$ fm$^{-3}$ where the $\alpha$ 
	particle is dissolved, is indicated.
	}
 \label{Fig:neu} 
 \end{figure}

\begin{figure}[hb] 
	\includegraphics[width=0.8\textwidth]{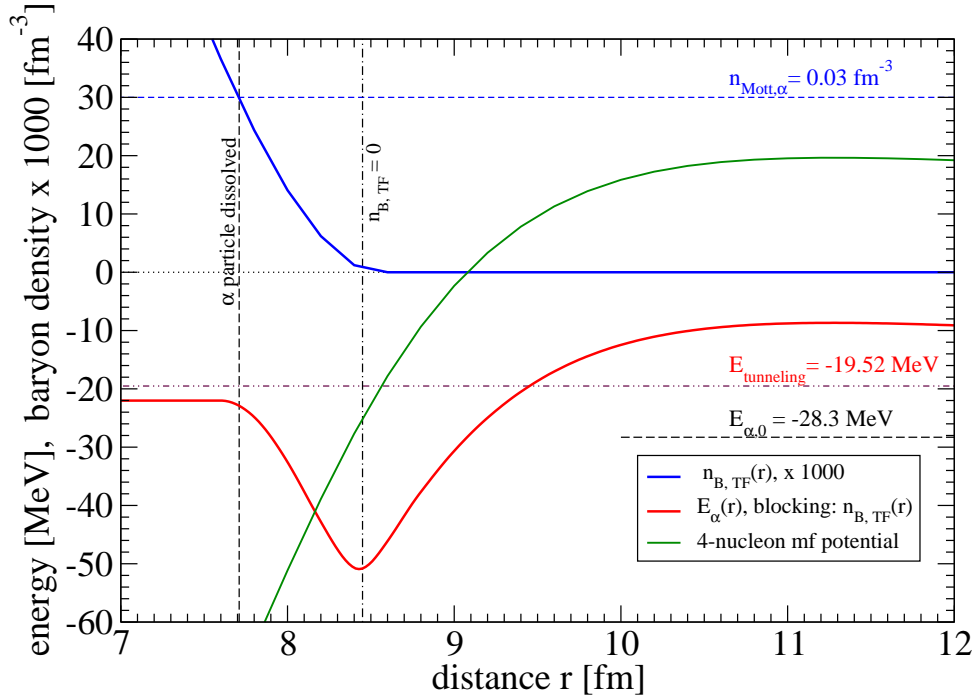}
	\caption{Insertion of Fig. \ref{Fig:neu}
	}
 \label{Fig:neu1} 
 \end{figure}

The effective c.o.m. potential $W({\bf R})$ is shown in Fig. \ref{Fig:neu}. 
At large distances, only the bound state energy of the free $\alpha$ particle remains, 
$\lim_{R \to \infty}W( R)=E^{(0)}_\alpha=-B^{(0)}_\alpha=-28.3$ MeV. 
For finite distances $R > 14$ fm, the Coulomb repulsion between the $\alpha$ particle and the lead core dominates the effective potential.
Below $R \approx 14$ fm,
the mean-field ($4 V^{\text{mf}}({\bf R})$) of the lead core becomes relevant tempting to attract the $\alpha$ particle. 
At distances below the Coulomb barrier, the intrinsic four-nucleon energy shifts strongly down.
As soon as the core nucleons have a finite density (within the Thomas-Fermi model at $R \approx  8.46$ fm), 
the blocking of the $\alpha$ particle acts.
The shift $W^{\rm Pauli}({\bf R})$ reduces the binding energy at distances $R$ 
where the densities of the $\alpha$ particle and the core nucleus overlap.
The bound state disappears if the baryon density $n_B$ approaches the value $n_{B,\text{cluster}}=0.0292$ fm$^{-3}$ what happens at 
$R_{\text{cluster}} \approx$ 7.72 fm. 
Then, the four-nucleon
system is at the Fermi energy $4 E_{\rm Fermi} \approx 22.014$ MeV, Eq. (\ref{4Fermi}), that is the edge of the quasi-particle continuum. 
At higher densities, the solution of the four nucleon problem is given by the single-nucleon shell states, 
and the empty states above the Fermi energy $E_{{\rm Fermi},\tau}=\mu_\tau$ are occupied by the added four nucleons 
on top of the $^{208}$Pb core.
An interesting result is the occurrence of a ``pocket'' near $R \approx 8.5$ fm 
in the effective $\alpha$-potential $W({\bf R})$. Details are shown in the insertion Fig.  \ref{Fig:neu1}

The calculations can be improved using a more detailed nucleon-nucleon interaction $V_{N-N}$ such as the Volkov force.
Furthermore, the intrinsic wave function can be improved within the variational approach similar to the treatment 
given in \cite{RSSN}, so that the $\alpha$-like cluster becomes more stable 
and the transition to the continuum states becomes smoother. The Pauli blocking is overestimated using $P=0$. 
A more appropriate expression for the Pauli blocking should also consider finite c.o.m. momenta for the 
bound, $\alpha$-like cluster state, see \cite{Ropke2011}.

The wave function $\Phi({\bf R})$ is calculated solving the corresponding Schr\"odinger equation (\ref{21c}).
The pocket of $W({R})$ shown in Fig. \ref{Fig:neu} is quite deep (-51.3 MeV at $R = 8.46$ fm) and  
a bound state at -32.47 MeV appears. The reason for the sharp minimum is the sharp disappearance of the nucleon density 
in the Thomas-Fermi model at the distance $R = 8.46$ fm where the mean-field potential $V^{\rm mf}(R)$ coincides with the chemical potential. 
More realistic nucleon densities of heavy nuclei show longer tails so that the 
Pauli blocking acts already at larger values of $R$. Nevertheless, we used the Thomas-Fermi model for our exploratory 
calculations because the physical background for the appearance of the potential pocket becomes more transparent. 
Future calculations have to improve this approximation so that the density distribution in the tail that is of relevance in our approach
is treated quantum mechanically. See App. \ref{app:3} for further discussions.

\section{Discussion and Conclusions}
\label{discussion}

The physics of cluster formation in homogeneous matter is reasonably well understood, however,
the numerical treatment is quite complex, see \cite{RSSN,Sogo}. 
There, the c.o.m. momentum is a good quantum number so that the separation 
into the c.o.m. motion and the relative motion is simple.
In addition to the formation of clusters which are modified by the surrounding matter, 
we have also the formation of quantum condensates such as pairing and quartetting. 

In the present work, we consider cluster formation in inhomogeneous nuclear systems, 
in particular $\alpha$-like clustering in heavy nuclei. We treated the particular situation where only a single 
$\alpha$ particle sits on top of a doubly magic nucleus which, thus, can be treated as a shell model core. 
In particular, we considered $^{212}$Po, that is one $\alpha$ on top of the $^{208}$Pb core. 
The $\alpha$ particle as a cluster plays a rather particular role among possible clusters. 
The physics of the deuteron is very different as explained in the main text. 
Heavier clusters may be treated with the fission-fusion scenario. 
At which mass number of the cluster the transition
from our present description to the latter one occurs, is not very clear. 
In any case the $\alpha$ particle is by itself a double magic nucleus (the lightest) and, 
therefore, very stable with its first excited state at $\sim $ 20 MeV. On the other hand, 
as we have shown in earlier works \cite{Ropke2011}, 
light clusters including the $\alpha$ particle are extremely sensitive to Pauli blocking from surrounding matter. 
Already at a fifth of saturation density, the $\alpha$ particle more or less suddenly becomes dissolved 
and gets mixed up with the surrounding Fermi gas. Translated to our $\alpha + ^{208}$Pb case, 
this means that an $\alpha$ approaching the Pb core stays a compact almost elementary particle until it feels 
the tail of the Pb density at around $\rho_0/5$. There, it quite suddenly dissolves and its four nucleons 
go over into single-particle shell model states with, eventually, pair correlations in the open shells on top of the $^{208}$Pb core. 
However, before its dissolution, the $\alpha$ particle already feels the attraction of the mean field of the core, 
so that one can understand the formation of a potential pocket at the surface of the Pb core. 

As we know, the description of a well defined cluster on top of a core nucleus is extremely difficult in an one center shell model description. 
Therefore, the main ingredient of this work is the introduction of the c.o.m. motion as a collective 
degree of freedom and an intrinsic motion that characterizes the cluster. To go beyond the single-quasiparticle approach, 
four-nucleon correlations are then described by an in-medium Schr\"odinger equation.
Besides the mean field, the crucial effect of the surrounding nuclear system is Pauli blocking 
as a consequence of antisymmetrization. As just explained, an $\alpha$-like bound state can exist 
in nuclear matter only at low densities,
$n_B \le n_{B,\text{cluster}}\approx 0.03$ fm$^{-3}$ and will be dissolved at higher densities into nearly 
free single-quasiparticle states forming the continuum of scattering states. It is clear that in a heavy 
nucleus only states near the Fermi energy can form an
$\alpha$-like cluster because only these single-particle states extend to the low-density regions at the surface of the nucleus.
Deeper mean field energy levels  are situated in the region of higher densities. 
There, the role of  cluster formation becomes irrelevant because of strong Pauli blocking.

The introduction of the c.o.m. motion and the intrinsic motion for clusters in nuclei, with full antisymmetrization of the 
nucleon wave function, was investigated within the THSR approach for light, low-density nuclei \cite{Boneu,THSR}. 
This gives a simple and adequate description of the properties of nuclei with cluster structure such as the Hoyle state.
We reconsidered the preformation of $\alpha$-like correlations within a generalized THSR approach that considers a fully 
antisymmetrized state of an $\alpha$-like cluster and the core nucleus. 
The c.o.m. motion of both constituents has to be treated in a consistent way. 
In contrast to a recent calculation for $^{20}$Ne  \cite{Bo,Bo2}, we replace the wave function of the doubly magic core nucleus by a 
shell model wave function. Furthermore we neglected the c.o.m. motion of 
the (very heavy) core nucleus because we treat a heavy system. However, for the 
non-localized $\alpha$ particle the c.o.m. motion is taken into account. 
After separation of the intrinsic motion within the $\alpha$ cluster,
an effective potential has been derived that describes the c.o.m. motion of the 
$\alpha$ cluster under the influence of Pauli blocking with the surrounding medium.

The approach presented in this work to include few-nucleon correlations, in particular bound states, 
is based on a first-principle approach to nuclear many-body systems. However, several approximations have been performed
to make the approach practicable and to work out the physical content. In particular, derivatives of the intrinsic wave function
$\varphi_4^{\text{intr}}({\bf s}_j,{\bf R})$ with respect to the c.o.m. co-ordinate $\bf R$ have been neglected.
For the nucleon-nucleon interaction $V_{N-N}$ a simple separable potential was taken, 
and Woods-Saxon like expressions have been used for the mean-field potential $V_\tau^{\rm mf}(r)$.
Furthermore, the effective c.o.m. potential $V_4^{\text{c.o.m.}}({\bf R})$ is taken in local approximation,
and instead of the correct single-particle states for a nucleus, the Thomas-Fermi (TF) model as a local-density approximation
was used. In general, the Pauli blocking as an exchange term leads to a non-local single particle potential. 
These approximations can be improved in more sophisticated future calculations.
The TF approximation for the c.o.m. motion of the $\alpha$ particle can be justified from the fact that 
before its abrupt dissolution, the $\alpha$ particle is still quite compact in extension, 
its radius having increased by only about 20 percent. Therefore, the extension of the $\alpha$ particle 
is never much larger than the surface width of the Pb core qualifying the TF approach as a reasonable lowest order approach.

The intrinsic energy, called $W(R)$, of the $\alpha$ particle, thus, becomes a function of the distance $R$ of the center 
of the core nucleus. It has two contributions. The effect of $W^{\text{Pauli}}({\bf R})$ is to reduce the attractive shift 
$W^{\rm ext}({\bf R})+E^{(0)}_\alpha$ of the four-nucleon cluster at distances $R$ 
where the densities of the $\alpha$ particle and the core nucleus overlap. 
It compensates the binding energy if the nucleon density $n_B( R)$ exceeds about 1/5 of the saturation density. 
This gives a microscopic derivation for the potential inferred by Delion and Liotta \cite{Delion}. 
The approach  \cite{Delion} considers a fixed position of the $\alpha$ particle as described by the pocket at a fixed position. 
This resembles the adiabatic approach in describing fission of $^{212}$Po into the two daughter nuclei. 
The approach presented here considers the non-localized  $\alpha$ particle where the c.o.m. motion is expressed 
by the wave function $ \Phi({\bf R}) $. The corresponding in-medium Schr\"odinger equations 
for the c.o.m motion and the intrinsic motion are derived within a quantum statistical approach.
This may serve also to further work out recent approaches using 
constrained Hartree-Fock calculations that have been performed for dilute 
nuclei showing a fragmentation of the mean field
and correspondingly the appearance of fragments \cite{Schuck13}, in particular to implement the c.o.m. motion. 

Let us discuss the relation of our present study of $^{212}$Po with respect to the similar situation of $^{20}$Ne 
which has been treated extensively already 40 years ago with the Resonanting Group Method (RGM), see Matsuse et al. \cite{Matsuse1} 
and also recently with the THSR wave function \cite{Bo,Bo2,Boneu}. In both cases one considers an $\alpha$ particle on top of a doubly magic core. 
In the case of $^{20}$Ne the core $^{16}$O is light and its c.o.m. motion must be treated correctly. 
This is done with the RGM as well as with THSR approaches. 
However, in the case of $^{212}$Po the $^{208}$Pb core is too massive for an application of those methods for technical reasons. 
On the other hand, this allows to treat the $^{208}$Pb core as infinitely heavy and then the corresponding treatment 
boils down to a four nucleon TDA equation as discussed earlier in the text. 
It is interesting to see that the effective $\alpha$ particle - core potentials for $^{20}$Ne and $^{212}$Po show some similarity. 
In both cases they become strongly attractive inside the Coulomb barrier, see e.g. Fig. 5 in \cite{Matsuse1} and Figs. 3,4 in present work. 
It would be interesting to also analyse the THSR approach in this respect.

A rigorous separation of the c.o.m. motion and the antisymmetrization can be made using Gaussian functions 
for the internal cluster wave functions as well as for the relative c.o.m. motion. 
This has been shown in several papers related to the THSR approach \cite{Funaki2011,Yamada2012,Yamada2011}. In particular, 
let us discuss the relation of our present treatment with the case of $^{20}$Ne consisting of $^{16}$O and an $\alpha$ cluster \cite{Bo}. 
Contrary to the latter case, we here supposed that the big cluster is infinitely heavy, 
so that we can represent it as shell model nucleus with a fixed c.o.m. position at $R=0$ from where all coordinates are measured. 
The antisymmetrization of the total wave function which we had in the case of 
$^{20}$Ne, is then here replaced by the Pauli blocking factors. This means that the THSR approach has the advantage that
the  $\alpha$ particle is treated in a correlated medium in contrast to the single particle, uncorrelated Pauli blocking term
($\Theta$ function in momentum space) considered in this work. A cluster-mean field approach \cite{RSM} would improve that. The extension of the original THSR
approach to heavy nuclei is numerically not feasible at present. On the other hand, in the double-magic $^{208}$Pb core nucleus
the $\alpha$-like correlations are not strong so that a shell model approach is reasonable.
Nevertheless, a comparison of the results obtained using the THSR ansatz with the approach given in our work if applied to light nuclei 
such as $^{20}$Ne would be of interest, as well as with former RGM calculations \cite{Matsuse} and recent investigations  \cite{Horiuchi}. 
The intrinsic wave function of the $\alpha$ particle has the same meaning as in $^{20}$Ne case. 
The c.o.m. wave function $\Phi({\bf R})$ plays the role of the relative wave function in the $^{20}$Ne case. 
The difficult point is the Pauli blocking factor which is a very non-local operator.

\begin{acknowledgments}
This work was initiated at the workshop on ``Clustering Aspects in Nuclei'', April 2013, at  the KITPC, Beijing, which was organized by Z. R.
The authors thank D. S. Delion for many interesting discussions at this workshop.

This work is supported by the National Natural Science Foundation
of China (Grants No. 11035001, No. 10975072, No. 10735010, No.
11375086, No. 11175085, No. 11235001, and No. 11120101005), by the
973 Program of China (Grants No. 2010CB327803 and No.
2013CB834400) and by the Project Funded by the Priority Academic
Program Development of Jiangsu Higher Education Institutions
(PAPD).

\end{acknowledgments}

\appendix

\section{Evaluation of the variational functional Eq. (\ref{varhom})}
\label{app}

We look for the minimum of the intrinsic energy, see Eq. (\ref{varhom}), of an $\alpha$-like cluster
\begin{eqnarray}
\label{varhomintr}
&&{\tilde W}^{\text{intr}}({\bf P})=\frac{\hbar^2}{2m} \int \frac{d^3 k}{(2 \pi)^3}\,
\frac{d^3 k_{12}}{(2 \pi)^3}\,\frac{d^3 k_{34}}{(2 \pi)^3}
\left[ k^2+2 k_{12}^2+2 k_{34}^2 \right]|{\tilde \varphi}^{\rm intr}({\bf k},
{\bf k}_{12},{\bf k}_{34},{\bf P})|^2 
 \\ &&
+\int \frac{d^3 k}{(2 \pi)^3}\,\frac{d^3 k_{12}}{(2 \pi)^3}\,\frac{d^3 k_{34}}
{(2 \pi)^3}
\,\frac{d^3 k'}{(2 \pi)^3}\,\frac{d^3 k'_{12}}{(2 \pi)^3}\,\frac{d^3 k'_{34}}
{(2 \pi)^3}
{\tilde \varphi}^{\rm intr,*}({\bf k},{\bf k}_{12},{\bf k}_{34},{\bf P}) 
V_4^{\rm intr}({\bf k}, {\bf k}_{12}, {\bf k}_{34},{\bf k}', 
{\bf k}'_{12}, {\bf k}'_{34},{\bf P})
{\tilde \varphi}^{\rm intr}({\bf k}',{\bf k}'_{12},{\bf k}'_{34},{\bf P})\nonumber.
\end{eqnarray}
The evaluation has been done for a special ansatz for the wave function, 
Eqs. (\ref{Gauss1}), (\ref{trial}) that contain a unique variational 
parameter $b$. The in-medium 4-particle interaction $V_4^{\rm intr}$ is given by (\ref{28aa}) with arbitrary ${\bf P}$.
The Pauli blocking is fulfilled by the ansatz (\ref{trial}) for the
wave function so that it must not considered any more.
For simplicity we consider only the c.o.m. momentum ${\bf P}=0$. (To discuss finite ${\bf P}$, a series expansion 
with respect to powers of $P$ can be performed.)  We have to transform from the single nucleon momenta ${\bf p}_i$
to Jacobi-Moshinsky momenta ${\bf k}_i$, Eq. (\ref{Jac2}).

To simplify the calculations we average the Fermi energies with respect to the isospin $\tau =n,p$ (symmetric matter), 
so that we perform the calculations for an excluded Fermi sphere 
$p_{\rm Fermi}=k_F=(3 \pi^2 n_B/2)^{1/3}$ with the total baryon density $n_B=n_n+n_p$.

The kinetic energy gives a 9-fold integral, the potential energy (after exploiting the $\delta$ functions) a 12-fold integral.
We use spherical coordinates where the integrals over the angles can be performed. By reason of isotropy, 
we can fix the direction of ${\bf k}$ and denote the $\cos \theta$ of the directions of ${\bf k}_{12}, {\bf k}_{34},
{\bf k}'_{12} $ relatively to ${\bf k}$ with $z_{12}, z_{34},z_{12}' $, respectively, i.e. $z_{12} = \cos({\bf k}_{12},{\bf k})$ etc. 
In Jacobi momenta, the expressions $F({\bf k},{\bf k}_{12},{\bf k}_{34})$ that have to be integrated have the form
\begin{equation}
 F({\bf k},{\bf k}_{12},{\bf k}_{34})\equiv F( k, k_{12},z_{12}, k_{34},z_{34})
\end{equation}
occurring for the norm or the kinetic energy, and with additional variables $k_{12}',z_{12}'$ for the potential energy.
For the functions $F$ considered here, the integral over $k$ is divided into two parts:
\begin{equation}
 \int d^3 k\,d^3 k_{12}\,d^3 k_{34} F({\bf k},{\bf k}_{12},{\bf k}_{34})=4 \pi \int_0^{2 k_F}dk G^<( k,{\bf k}_{12},{\bf k}_{34})
 +4 \pi \int_{2 k_F}^\infty dk G^>( k,{\bf k}_{12},{\bf k}_{34}).
\end{equation}
Next we consider the integral over ${\bf k}_{12}$. The excluded region in momentum space that is occupied by the Fermi sphere
leads to a restriction of the limits of the integrals over $z_{12} = \cos({\bf k}_{12},{\bf k})$. Geometrical considerations give
for $k\le 2 k_F$ the following limits where the Fermi sphere is touched,
\begin{eqnarray}
&&G^<( k,{\bf k}_{12},{\bf k}_{34})= \int d^3 k_{12} H^<(   k,k_{12},z_{12},{\bf k}_{34})\\
&&=2 \pi \left[ \int_{\sqrt{k_F^2-k^2/4}}^{k_F+k/2} k_{12}^2 dk_{12} 2 \int_{(k_F^2-k^2/4-k_{12}^2)/kk_{12}}^0 dz_{12}
H^<( k,k_{12},z_{12},{\bf k}_{34})
 +\int_{k_F+k/2}^\infty k_{12}^2 dk_{12} \int_{-1}^1 dz_{12}H^<(  k,k_{12},z_{12},{\bf k}_{34})\right]\nonumber,
\end{eqnarray}
and for $k\ge 2 k_F$
\begin{eqnarray}
&&G^>( k,{\bf k}_{12},{\bf k}_{34})= \int d^3 k_{12} H^>(  k,k_{12},z_{12},{\bf k}_{34})\\
&&=2 \pi \left[ \int_{0}^{k/2-k_F} k_{12}^2 dk_{12} \int_{-1}^1 dz_{12}H^>(  k,k_{12},z_{12},{\bf k}_{34})
+\int_{k/2+k_F}^\infty k_{12}^2 dk_{12} \int_{-1}^1 dz_{12}H^>(  k,k_{12},z_{12},{\bf k}_{34})\right.\nonumber \\ &&\left.+
  \int_{k/2-k_F}^{k/2+k_F} k_{12}^2 dk_{12} 2 \int_{(k_F^2-k^2/4-k_{12}^2)/kk_{12}}^0 dz_{12}H^>(  k,k_{12},z_{12},{\bf k}_{34})\,.
 \right]\nonumber
\end{eqnarray}
The remaining integrals are performed in the same way. For the special trial function (\ref{trial}), 
the integral over the angular part $z_{12}$
etc. can be performed analytically. The norm, the kinetic energy, and the potential energy are calculated as integral over $k$ 
after the relative momenta 
$k_{12},k_{34},k'_{12}$ have been integrated over. Thus, the 9 or 12 fold integrals are reduced to 3 or 4 fold integrals, respectively,
that can be handled. For a given density, that also determines the blocked phase space for 
the four-particle wave function, the trial wave function (\ref{trial}) contains the parameter $b$ which describes how fast the wave 
function is decreasing with increasing single-particle momentum. For a similar evaluation of multiple integrals see also \cite{Sogo}.

With this variational ansatz, the minimum of the energy is determined for the optimal $b$ parameter for each density. 
Results are given in Section \ref{intrinsichom}. To improve the variational solution of the wave equation (\ref{10}) for the intrinsic motion,
the class of functions (\ref{trial}) can be extended.

\section{Local approximation for the Pauli blocking term}
\label{app:2}

As an example, we consider the term $\langle {\bf r}_1 {\bf r}_2 |f_1(\varepsilon_{n_1})|{\bf r}''_1 {\bf r}''_2 \rangle
=\langle {\bf r}_1| f_1(\varepsilon_{n_1})|{\bf r}''_1 \rangle\delta({\bf r}''_2-{\bf r}_2)$ occurring in $B$.
We transform into a "mixed" (Wigner) representation,
\begin{equation}
\langle {\bf r}_1| f_1(E_{n_1})|{\bf r}''_1 \rangle =\int \frac{d^3 p_{1}}
{(2 \pi)^3} \,
{\rm e}^{i {\bf p}_1 \cdot ({\bf r}_1-{\bf r}''_1)}
f^{\rm Wigner}_1\left(\frac{{\bf r}_1+{\bf r}''_1}{2}, {\bf p}_1\right)\,.
\end{equation}
The occupation of the phase space is given by the quasi-particle wave functions $\psi_n({\bf r})$ 
(we take $\frac{{\bf r}_1+{\bf r}''_1}{2}={\bf R}_1$)
\begin{equation}
 f^{\rm Wigner}_1\left({\bf R}_1, {\bf p}_1\right)=\int d^3 s_1 
{\rm e}^{-i {\bf p}_1 \cdot {\bf s}_1}
 \sum_n^{\rm occupied}\psi^*_n({\bf R}_1-\frac{{\bf s}_1}{2})\psi_n({\bf R}_1+
\frac{{\bf s}_1}{2})\,.
\end{equation}
Within the Thomas-Fermi model that corresponds to a local density approximation (or rather 'local momentum approxiamtion'), we have
\begin{equation}
\label{WignerR}
 f^{\rm Wigner}_1\left({\bf R}_1, {\bf p}_1\right)\approx \Theta\left(E_{\rm Fermi}
[n_B({\bf R}_1)]-\frac{ p_1^2}{2 m}\right),
\end{equation}
see also Eq. (\ref{blocking}).
The phase space occupation is determined by the Fermi energy $E_{\rm Fermi}(n_B)=(\hbar^2/2m) (3 \pi^2n_B/2)^{2/3}$ 
where we consider for simplicity the symmetric case $n_n=n_p=n_B/2$ as in Sec. \ref{homogen}. 
Now, the baryon density $n_B({\bf R}_1)$ depends on the position ${\bf R}_1$.

Again we introduce Jacobi coordinates to extract the c.o.m. motion as collective degree of freedom so that
\begin{eqnarray}
&&\langle {\bf R},{\bf s},{\bf s}_{12},{\bf s}_{34}| f_1(\varepsilon_{n_1})|{\bf R}'',{\bf s}'',{\bf s}''_{12},{\bf s}''_{34} \rangle 
= \int \frac{d^3 p_{1}}{(2 \pi)^3} \,{\rm e}^{i {\bf p}_1 \cdot ({\bf s}_{12}-{\bf s}''_{12})}
f^{\rm Wigner}_1\left({\bf R}+\frac{{\bf s}+{\bf s}''_{12}}{2}, {\bf p}_1\right) 
\nonumber \\ && \times\delta({\bf s}''_{34}-{\bf s}_{34})
\delta({\bf s}''-{\bf s}-2{\bf R}''+2{\bf R})\delta({\bf s}''_{12}-{\bf s}_{12}-4{\bf R}''+4{\bf R})\,.
\end{eqnarray}

To evaluate the contribution of $V^{\rm intr}_4$ to the intrinsic energy we use 
a ``mixed'' representation where the intrinsic motion is given in momentum representation,
\begin{equation}
\label{phiks}
 \tilde \varphi_4^{\text{intr}}({\bf k},{\bf k}_{12},{\bf k}_{34},{\bf R})=\int d^3 s\, d^3 s_{12} \, d^3 s_{34} 
 {\rm e}^{-i {\bf k} \cdot {\bf s}-i{\bf k}_{12} \cdot {\bf s}_{12}-i{\bf k}_{34} \cdot {\bf s}_{34}}
 \varphi_4^{\text{intr}}({\bf s},{\bf s}_{12},{\bf s}_{34},{\bf R})\,.
\end{equation}
To evaluate the Pauli blocking contribution to the effective c.o.m. 
potential $ W({\bf R},{\bf R}')  $, 
Eqs. (\ref{9bb}), (\ref{W4R}), (\ref{11c}), we have to average over the 
intrinsic motion. We consider here only one of the different terms 
(the others follow analogously).  We give the general expression 
which is complicated but will immediately be reduced to a more tractable 
form below in (\ref{F1}).
 \begin{eqnarray}
\label{9bbb}
&&F_1({\bf R},{\bf R}')=\int d^9s_j\,d^9s'_j\,d^9s''_j\,d^3 R''\,
\varphi_4^{\text{intr},*}({\bf s},{\bf s}_{12},{\bf s}_{34},{\bf R}) 
\langle {\bf R},{\bf s},{\bf s}_{12},{\bf s}_{34}| f_1(\varepsilon_{n_1})|
{\bf R}'',{\bf s}'',{\bf s}''_{12},{\bf s}''_{34} \rangle 
\nonumber \\ &&
\times \langle {\bf s}'', {\bf s}''_{12}, {\bf s}''_{34} |V_{N-N}|{\bf s}', 
{\bf s}'_{12}, {\bf s}'_{34} \rangle \delta({\bf R}''-{\bf R}') 
\varphi_4^{\text{intr}}({\bf s}',{\bf s}'_{12},{\bf s}'_{34},{\bf R}')\,.
\end{eqnarray}

\begin{eqnarray}
\label{RR"}
&&F_1({\bf R},{\bf R}')=\int d^3 s\, d^3 s_{12} \,   d^3 s''_{12}   \,
d^3 k\, d^3 k_{12} \, d^3 k_{34}  d^3 k'\, d^3 k'_{12} \,  d^3 k''_{12}\, \frac{d^3 p_{1}}{(2 \pi)^{21}}
\tilde \varphi_4^{\text{intr},*}({\bf k},{\bf k}_{12},{\bf k}_{34},{\bf R})
{\rm e}^{i {\bf k} \cdot {\bf s}+i{\bf k}_{12} \cdot {\bf s}_{12}}
\nonumber \\ && \times
{\rm e}^{i {\bf p}_1 \cdot ({\bf s}_{12}-{\bf s}''_{12})}
f^{\rm Wigner}_1\left({\bf R}+\frac{{\bf s}+{\bf s}''_{12}}{2}, {\bf p}_1\right) 
{\rm e}^{i {\bf k}' \cdot ({\bf s}-{\bf s}_{12}/2+{\bf s}''_{12}/2)+i{\bf k}''_{12} \cdot {\bf s}''_{12}}
V_{N-N}({\bf k}''_{12};{\bf k}'_{12})
\nonumber \\ && \times
\varphi^{\text{intr}}\left({\bf k}',{\bf k}'_{12},{\bf k}_{34},{\bf R}-\frac{{\bf s}_{12}-{\bf s}''_{12}}{4}\right) 
\delta\left({\bf R}'-{\bf R}+\frac{{\bf s}_{12}-{\bf s}''_{12}}{4}\right)\,.
\end{eqnarray}
As expected, expression (\ref{9bbb}) is not local in ${\bf R}$. The Wigner function limits the ${\bf p}_1$ integral
as $\int_0^{p_{\rm Fermi}[n_B({\bf R}+\frac{{\bf s}+{\bf s}''_{12}}{2})]}\frac{d^3 p_{1}}
{(2 \pi)^3}$. 
We can expand near ${\bf R}$ so that additional terms near the Fermi surface are neglected. Also the wave function 
${\tilde \varphi}^{\text{intr}}({\bf k}',{\bf k}'_{12},{\bf k}_{34},{\bf R}-
\frac{{\bf s}_{12}-{\bf s}''_{12}}{2})$
can be expanded near ${\bf R}$. Neglecting higher order contributions we have
\begin{eqnarray}
&&F_1({\bf R},{\bf R}')\approx \int d^3 s\, d^3 s_{12} \,   d^3 s''_{12}   \,
d^3 k\, d^3 k_{12} \, d^3 k_{34}  d^3 k'\, d^3 k'_{12} \,  d^3 k''_{12}\, 
\frac{d^3 p_{1}}{(2 \pi)^{21}}
\tilde \varphi_4^{\text{intr},*}({\bf k},{\bf k}_{12},{\bf k}_{34},{\bf R}){\rm e}^{i {\bf k} \cdot {\bf s}+i{\bf k}_{12} \cdot {\bf s}_{12}}
\nonumber \\ && \times
{\rm e}^{i {\bf p}_1 \cdot ({\bf s}_{12}-{\bf s}''_{12})}f^{\rm Wigner}_1\left({\bf R}, {\bf p}_1\right) 
{\rm e}^{i {\bf k}' \cdot ({\bf s}-{\bf s}_{12}/2+{\bf s}''_{12}/2)+i{\bf k}''_{12} \cdot {\bf s}''_{12}}
V_{N-N}({\bf k}''_{12};{\bf k}'_{12})
{\tilde \varphi}_4^{\text{intr}}({\bf k}',{\bf k}'_{12},{\bf k}_{34},{\bf R})\delta\left({\bf R}'-{\bf R}\right)
\end{eqnarray}
that is diagonal in ${\bf R}$ space. Higher order terms are connected with 
intrinsic coordinates ${\bf s}_j$ and are averaged
out with the intrinsic wave function. The local approximation contains 
leading terms but can be improved in a systematic way.

Now we can integrate over the intrinsic coordinates ${\bf s}_j$ and obtain
\begin{eqnarray}
\label{F1}
&&F_1({\bf R},{\bf R}')= \int d^3 k\, d^3 k_{12} \, d^3 k_{34}  \frac{ d^3 k'_{12}}{(2 \pi)^{12}} \, 
\nonumber \\ && \times
\tilde \varphi_4^{\text{intr},*}({\bf k},{\bf k}_{12},{\bf k}_{34},{\bf R})
f^{\rm Wigner}_1\left({\bf R}, {\bf k}+{\bf k}_{12}\right) 
V_{N-N}({\bf k}_{12};{\bf k}'_{12})
\varphi_4^{\text{intr}}({\bf k}',{\bf k}'_{12},{\bf k}_{34},{\bf R})
\delta\left({\bf R}'-{\bf R}\right)\,.
\end{eqnarray}

The integral over ${\bf k}_{1}= {\bf k}+{\bf k}_{12}$ is restricted to the Fermi sphere, 
$k_1 \le k_{\rm Fermi,\tau}[n({\bf R}+\frac{{\bf s}_{1}+{\bf s}'_{1}}{4})]$.
We can expand with respect to $\frac{{\bf s}_{1}+{\bf s}'_{1}}{4}$
so that the integral over the Fermi sphere gets additional contributions at $E_{{\rm Fermi},\tau}$.
In zeroth order, we have only the Fermi energy at the c.o.m. position ${\bf R}$, but this can be improved
taking the terms  $\frac{{\bf s}_{1}+{\bf s}'_{1}}{4}$ into account.

\section{Improved density profile}
\label{app:3}

The Thomas-Fermi model gives a rather sharp pocket so that the $\alpha$-particle is well formed at the surface of the core nucleus.
However, then the core nucleus may form further clusters so that the single-nucleon Thomas-Fermi approximation is not consistent.
The Thomas-Fermi model is a quasi classical approach that cannot describe the behavior of the tails of the density distribution 
that are of relevance for the cluster formation.
Thus, the nucleon density abruptly disappears at the value of the radius where the Fermi energy 
coincides with the potential energy. Real density distributions are more smooth and show long tails also in the region where 
the potential energy is larger than the Fermi energy due to quantum tunneling. A shell model calculation would give a better description of that region.

\begin{figure}[hb] 
	\includegraphics[width=0.8\textwidth]{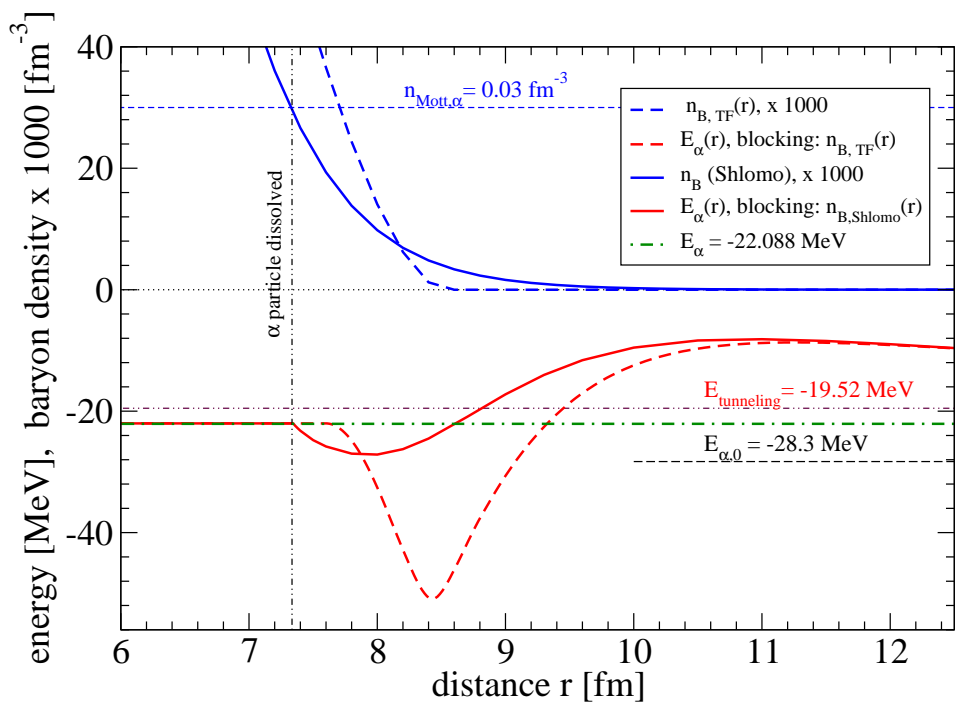}
	\caption{Nucleon density $n_B( r)$ (blue, full) according to Shlomo \cite{Shlomo} compared with the Thomas-Fermi approach 
	(blue, dashed).
	The local effective potential  $W({\bf R})$ (\ref{Vcm}) (red, dashed) according to the Thomas-Fermi approach
	is compared with an effective potential (red, full)  that describes the Pauli blocking of the long-ranged density tails at the 
	surface of the lead core. The ground-state energy level (green, dash-dotted) is also shown.}
 \label{Fig:neuShlomo} 
 \end{figure}

Since we are interested in the region where the nucleon density is low, we 
discuss here the consequences of long-range tails of the density. We use the  nucleon density profile according to Shlomo \cite{Shlomo}
$n_B( R)=0.17 \{1+\exp [-(R-6.4914)/0.54]\}^{-1}$ (units in fm) for the lead core. 
This density profile is also shown in Fig. \ref{Fig:neuShlomo}.
The region of finite density, where Pauli blocking occurs, is extended to higher distances, above the value 8.46 fm obtained 
in the Thomas-Fermi model. The critical density where the   $\alpha$-particle is dissolved occurs at the distance $R=7.3416$ fm 
that is smaller than the value given by the Thomas-Fermi model. These considerations are based on the given density profile,
and instead of the theoretical estimates also experimental density distributions can be used. 
In a more detailed approach, different density profiles for neutrons and protons can be considered.

Now, we have to introduce the mean-field potential of the core nucleons. If the position of the Fermi energy 
(-22.014 MeV) remains unchanged, 
the Woods-Saxon potential yields too much density at  $R=7.3416$ fm. Since now the density profile is inferred due to the Shlomo approach, 
we can adapt the Woods-Saxon potential correspondingly
so that the value of the critical density is reproduced within the Thomas-Fermi approximation. 
For our present estimation, the contribution of the neutron potential  (\ref{WSn}) 
to the four-nucleon potential was reduced by the factor 0.585. Then the condition for the disappearance of the  $\alpha$-particle
at the critical radius $R_{\rm crit}$ is correctly implemented.
If we solve the c.o.m. Schr\"odinger equation with the effective c.o.m. potential we find the bound state energy at -22.088 MeV.

For illustration, 
in Fig. \ref{Fig:neuShlomo} the nucleon density for the $^{208}$Pb core according to Shlomo \cite{Shlomo} is shown.
It is clearly seen that the tail of the nucleon density $n_B$ extends to larger values of $R$. Therefore,
the pocket becomes shallow, and the bound state energy of the four-nucleon ($\alpha$-like) bound state 
becomes less negative. In an exploratory calculation where the Pauli blocking is calculated with the nucleon density 
profile according to Shlomo \cite{Shlomo}, the minimum of the pocket is -27.21 MeV at $R = 7.93$ fm. The corresponding solution 
of the c.o.m. Schr\"odinger equation (\ref{21c}) yields a bound state energy at -22.088 MeV in better agreement with the 
empirical value -19.52 MeV. More systematic calculations based on shell model states instead of the Thomas-Fermi model 
will provide us with more accurate results solving the c.o.m. motion of the $\alpha$-like cluster 
on top of a heavy core nucleus.

The possibility to describe four-particle correlations and preformed $\alpha$-like clusters near the surface 
can provide us with a theoretical tool to attack the cluster structure of nuclei like $^{212}$Po where until now only semi-empirical 
approaches are known to determine the $\alpha$ decay. $\alpha$-like correlations can survive in nuclear matter only up to densities
$n_B \le 0.03$ fm$^{-3}$, i.e. in the outer region of the nucleus. Preformation of $\alpha$ clusters and the $\alpha$-like content 
of the four-nucleon wave function can be treated within the approach given here. For this, we consider the four-nucleon wave function 
$\Psi({\bf R}, {\bf s}_j)$ (\ref{4}). The intrinsic part $\varphi^{\rm intr}$ is normalized for each ${\bf R}$, and the c.o.m. part
$\Phi({\bf R})$ that follows from the solution of a wave equation with the effective c.o.m. potential $W$, is normalized as well.
For an estimation, we assume that in the region where the $\alpha$-like cluster may exist the overlap of the intrinsic wave function 
with the free $\alpha$ intrinsic wave function is equal to one, and it is zero in the remaining part where intrinsic wave function of  the four-nucleon
system is given as product of single-nucleon states. 
We integrate over the space $R \ge R_{\rm cluster}$ to find the amount of $\alpha$ clustering,
\begin{equation}
 S=\int_0^\infty  d^3R |\Phi({\bf R})|^2 \Theta \left[n_B^{\rm critical}-n_B({\bf R})\right] \approx 0.371
\end{equation}
where $R_{\rm cluster}$ denotes the radius where the baryon density has the 
critical value where $\alpha$-like clusters are destroyed because of Pauli blocking, $n_B(R_{\rm cluster})= n_B^{\rm critical}=0.03$ 
fm$^{-3}$. This result is in reasonable agreement with other estimations, see \cite{Varga,Delionbuch}. Note that the intrinsic state remains $\alpha$-like as far as the change of the width parameter $b$ is small. 

The results given here should be improved by systematic shell model calculations. We emphasize that our treatment, 
worked out with some approximations to allow for exploratory calculations, gives the possibility to improve the
approximations, in particular the local density approximation, the neglect of the gradient terms of the intrinsic wave function, 
the introduction of non-local interaction potentials. Furthermore, a systematic improvement of the Green functions approach
allows also to include correlations in the nuclear matter that is treated here as uncorrelated medium, in contrast to the THSR approach
that treats four-particle correlations coherently.

\end{document}